\documentclass[12pt]{article}

\pdfoutput=1

\usepackage{ifpdf}
\ifpdf
\usepackage{graphicx,color}
\usepackage{hyperref}
\else
\usepackage[dvipdfmx]{graphicx,color}
\usepackage[dvipdfmx]{hyperref}
\fi
\usepackage{amssymb,amsfonts,amsmath,cancel,cite,multirow}
\usepackage{tikz}
\usepackage[capitalise]{cleveref}
\usepackage[margin=1in]{geometry}

\newcommand{\eqs}[1]{\begin{equation}\begin{split} #1 \end{split}\end{equation}}
\newcommand{\tikzcircle}[2][black,fill=black]{\tikz[baseline=-0.5ex]\draw[#1,radius=#2] (0,0) circle ;}%

\begin{document}

\begin{titlepage}

\begin{flushright}
\end{flushright}

\vskip 1.35cm
\begin{center}

{\large
\textbf{
    Quantum Theory of Dark Matter Scattering
}}
\vskip 1.2cm

Ayuki Kamada$^{a}$,
Takumi Kuwahara$^{b}$, 
and Ami Patel$^{a}$

\vskip 0.4cm

\textit{$^a$
    Institute of Theoretical Physics, Faculty of Physics, University of Warsaw, ul. Pasteura 5, PL-02-093 Warsaw, Poland
}

\textit{$^b$
    Center for High Energy Physics, Peking University, Beijing 100871, China
}

\vskip 1.5cm

\begin{abstract}
  Dark matter self-scattering is one of key ingredients for small-scale structure of the Universe, while dark matter annihilation is important for the indirect measurements. 
  There is a strong correlation between the velocity-dependent self-scattering cross section and the Sommerfeld enhancement factor for the dark matter annihilation cross section.
  In this study, we formulate a direct relation between them by the use of Watson's (initial state/final state) theorem and Omn\`es solution, and our formulation reproduces the Sommerfeld enhancement factor, which directly computed by solving the Schr\"odinger equation, from the scattering phase shift.
\end{abstract}

\end{center}
\end{titlepage}

\section{Introduction}

Interaction is one of the most important properties of particles. 
The interaction of dark matter (DM) is still unknown even though the existence of DM has been firmly confirmed by astrophysical observations. 
In this study, we focus on two kinds of interactions of DM: annihilation and self-scattering processes.

Recently, there has been growing interests in DM self-interaction.
We can prove the self-interaction of DM in cosmological observations though it is extremely challenging to find the self-interaction of DM in terrestrial experiments.
DM self-scattering may explain the small-scale structure problems of the Universe~\cite{Spergel:1999mh} (see Ref.~\cite{Tulin:2017ara} a review).
It has been extensively studied that these problems may be resolved even within the collisionless cold DM paradigm once we take into account baryon feedback processes in simulations.
However, it is still under debate to what extent the baryon feedback processes accommodate the problems, and hence it is intriguing possibility that DM has self-interaction.
Combining observations of DM halos in a wide range of masses may indicate the velocity dependence of the DM self-scattering cross section~\cite{Kaplinghat:2015aga}.
The self-scattering cross section per mass is preferred to be $\sigma/m \sim 0.1 \, \mathrm{cm^2/g}$ at the galaxy clusters, where the collision velocity is of $\mathcal{O}(10^3) \, \mathrm{km/s}$, in order to explain the core formation of the galaxy clusters~\cite{Newman:2012nw} and to avoid the constraints of the merging clusters~\cite{Randall:2008ppe,Harvey:2018uwf}.
As for the galaxy scale, where the velocity is of $\mathcal{O}(10^2) \, \mathrm{km/s}$, $\sigma/m \sim 1 \, \mathrm{cm^2/g}$ is preferred to explain the galactic rotation curve~\cite{Kamada:2016euw,Ren:2018jpt}. 
The DM self-scattering in the Milky Way (MW) dwarf spheroidal galaxies, where a typical velocity is $v \sim 30\,\mathrm{km/s}$, is still under debate. 
The preferred scattering cross section per mass is widely distributed as $0.3 \text{--} 40 \,\mathrm{cm^2/g}$~\cite{Valli:2017ktb}.
The expected cross section can be enlarged if DM halos are in the core-contraction phase: $30 \text{--} 120 \,\mathrm{cm^2/g}$~\cite{Correa:2020qam}.
On the other hand, the ultra-faint dwarf galaxies, which are considered to be more DM-dominated, provide constraints on the cross section to be $\sigma/m \lesssim 0.1 \, \mathrm{cm^2/g}$~\cite{Hayashi:2020syu}.
If the preferred scattering cross section per mass is of $\mathcal{O}(1 \text{--} 10^2) \,\mathrm{cm^2/g}$ for the typical velocity of $v \sim 30\,\mathrm{km/s}$, DM models with a light mediator are one of the simplest models to explain the velocity-dependence of the self-scattering cross section.

Meanwhile, the DM annihilation into the standard-model (SM) particles (or the mediator particles) is important for determining the DM abundance today in some scenarios with the freeze-out mechanism. 
The annihilation rate today is also important for the indirect searches for DM.
The annihilation cross sections for these processes can significantly differ due to different velocities even if the microscopic interactions controlling these processes are same. 
A typical velocity of DM is $v/c \sim \mathcal{O}(0.1)$ at freeze-out, while the maximum velocity of DM is $v/c \sim \mathcal{O}(10^{-2})$ at the cosmic structures. 
When the DM particle couples to a light force mediator, the annihilation cross section can be significantly enhanced in the low velocity regime~\cite{Hisano:2002fk,Hisano:2003ec,Hisano:2004ds,Cirelli:2007xd,Arkani-Hamed:2008hhe}, which is known as the Sommerfeld enhancement~\cite{Sommerfeld:1931}.
The indirect measurements of DM put strong constraints on the DM annihilation into the SM particles, such as $e^\pm$ and $\gamma$.
On the contrary, the velocity-dependent annihilation cross section has been often utilized to explain the observed cosmic-ray excesses for a decade after the positron excess (such as PAMELA~\cite{PAMELA:2008gwm}, Fermi-LAT~\cite{Fermi-LAT:2011baq}, and AMS-02~\cite{AMS:2019rhg}) was reported~\cite{Arkani-Hamed:2008hhe}.

In the context of the self-interacting DM (SIDM), the presence of the light mediator leads not only to the velocity-dependent self-scattering but also to a large annihilation cross section. 
When the Sommerfeld-enhanced DM annihilation ends up with the energy injection into the electromagnetic channels, the indirect measurements put strong constraints on the annihilation cross section~\cite{Bringmann:2016din}.
The Sommerfeld enhancement factor can also be sizable for the SIDM model providing the self-scattering cross section that almost saturates the Unitarity bound, as pointed out in Ref.~\cite{Kamada:2020buc}.
There can be a connection through quantum mechanical properties between the annihilation process and the scattering process. 
The Watson's (final state/initial state) theorem~\cite{Watson:1952ji} states that the discontinuity of a weak-coupling annihilation amplitude is given by the phase shift of the relevant scattering process.
The annihilation amplitude is given by so-called Omn\`es solution as far as the scattering process is in the elastic regime. 
In this study, we apply the Omn\`es solution to a system that contains the DM particle $\chi$ and the light mediator particle $\phi$.
The amplitude for DM annihilation into $\phi$ particles is related with the phase shift of the self-scattering process via the Watson's theorem.

The organization of this paper is as follows.
In the next section, we introduce the standard methods to compute the phase shift for the DM self-scattering and the Sommerfeld enhancement factor for the DM annihilation under the attractive Yukawa potential, and we confirm that there is a strong correlation between the resonant self-scattering and the large Sommerfeld enhancement factor. 
In \cref{sec:Watson}, we introduce the Watson's final-state/initial-state theorem and the Omn\`es solution, which relate the behavior of the scattering phase shift and the annihilation amplitude. 
We also provide a new computational method for the Sommerfeld enhancement factor via the Omn\`es solution. 
One of key ingredients for our formulation is the number of bound states and corresponding poles of the amplitudes.
We summarize the phase shift and the bound states by introducing the effective range theory in \cref{sec:ERT}. 
We can find the pole of the annihilation amplitude corresponding to the shallow bound state by the use of the effective range theory. 
In \cref{sec:Yukawa}, we demonstrate the agreement of the Sommerfeld enhancement factors in two methods.
Even though we focus on the Yukawa potential in the main text, we also discuss the attractive Hulth\'en potential in \cref{app:Hulthen}.
\cref{sec:conclusion} is devoted to conclude our study.

\section{Self-Scattering and Annihilation \label{sec:corr}}

In this section, we introduce the standard computational methods for the phase shift of the DM self-scattering and for the Sommerfeld enhancement factor of the DM annihilation in the presence of the long-range force.
We consider a system composed of DM particle $\chi$ (with the mass of $m_\chi$) and mediator particle $\phi$ (with the mass of $m_\phi$) interacting through the attractive Yukawa potential%
\footnote{
    We also discuss the Hulth\'en potential, which approximates the Yukawa potential and gives an analytical expression, in \cref{app:Hulthen}.
}:
\eqs{
    V(r) = - \frac{\alpha e^{- m_\phi r}}{r} \,.
}
Here, $r$ denotes the relative distance between two DM particles, and $\alpha$ is a fine-structure constant of the long-range force.
We define the dimensionless parameters, $a \equiv v/2 \alpha$ and $b \equiv \alpha m_\chi/m_\phi$ with $v$ being the relative velocity.

First, we discuss the DM self-scattering, $\chi\chi \to \chi\chi$, described by the Yukawa potential.
For DM self-scattering with low velocity mediated by the light particle, neither the Born approximation ($b < 1$, for which the cross section is computed perturbatively) nor the classical approximation ($a b > 1$, for which a classical cross section describes the DM scattering) is valid. 
The non-perturbative effect by a multiple exchange of the light mediator becomes important for the parameter space, $b > 1$ and $a < 1$. 
Since we do not have an approximate formula for the cross section $\sigma$ for such parameter space, we must solve the Schr\"odinger equation for the DM two-body system,
\eqs{
    \left[ - \frac{1}{2\mu} \nabla^2 + V(r) \right] \psi_k(x) 
    = E \psi_k(x)\,, 
}
with the partial-wave analysis in order to take into account the non-perturbative effect.
Here, $\mu = m_\chi/2$ is the reduced mass, $E = k^2/2 \mu$, and $k = \mu v$ is the relative momentum.
The wave-function has an asymptotic form at large $r$ as 
\eqs{
    \psi_k(x) \to e^{ikz} + f(k,\theta) \frac{e^{ikr}}{r} \,.
}
Here, $f(k,\theta)$ denotes the scattering amplitude, and the differential cross section for the scattering process is given by $d\sigma = |f(k,\theta)|^2 d \Omega$.

The scattering state is decomposed into its partial-wave contributions (with multipole $\ell$) as
\eqs{
    \psi_k(x) = \sum_\ell \frac{1}{k} e^{\frac{i}{2} \ell \pi + i \delta_\ell} (2 \ell + 1) R_{k,\ell} (r) P_\ell(\cos\theta) \,,
}
and the one-dimensional Schr\"odinger equation for the radial wave-function $R_{k,\ell}$ is written as 
\eqs{
    \frac{1}{r^2} \frac{d}{d r} \left( r^2 \frac{d R_{k,\ell}}{d r} \right) + \left( k^2 - \frac{\ell(\ell+1)}{r^2} - 2 \mu V(r) \right) R_{k,\ell} = 0 \,.
    \label{eq:rad_Sch}
} 
Since the potential term rapidly vanishes as $r \to \infty$, the asymptotic form of the radial wave-function at large $r$ is 
\eqs{
    R_{k,\ell}(r) \to \frac{1}{r} \sin\left( k r - \frac12 \ell \pi + \delta_\ell(k) \right) \,,
    \label{eq:asymptotic_Rkl}
}
where the scattering phase shift $\delta_\ell$ is real. 
We choose the normalization of $\delta_\ell$ to be $\delta_\ell (k) \to 0$ as $k^2 \to \infty$ since the potential term is negligible at high velocity.
From the asymptotic form of $\psi_k(x) - e^{i k r \cos\theta}$, we obtain the partial-wave decomposition of the scattering amplitude as follows:
\eqs{
    f(k,\theta) = \sum_\ell (2 \ell + 1) f_\ell(k) P_\ell(\cos\theta) \,, \qquad 
    f_\ell(k) \equiv \frac{e^{2 i \delta_\ell(k)} -1}{2ik} 
    = \frac{1}{k \cot \delta_\ell - i k} \,.
    \label{eq:sc_amp_decom}
}
The elastic scattering cross section is decomposed into its partial-wave contributions.
\eqs{
    \sigma = \sum_\ell \frac{4 \pi}{k^2} (2 \ell +1) \sin^2 \delta_\ell \,.
    \label{eq:crosssection}
}

The radial wave-function $R_{k,\ell}(r)$ for the two-body DM state also determines an important effect for the DM annihilation, $\chi\chi \to \phi\phi$, with low velocity, namely the Sommerfeld enhancement factor.
The annihilation occurs near $r = 0$, and the wave-function distorted from plane-wave near the origin significantly changes the size of the annihilation cross section.
The enhancement factor with multipole $\ell$, $S_\ell$, is given by the ratio of wave-functions with and without the potential at the origin~\cite{Iengo:2009ni,Cassel:2009wt} 
\eqs{
    S_\ell(k^2) 
    = \frac{\sigma_{\mathrm{ann},\ell} v}{(\sigma_{\mathrm{ann},\ell} v)_0} 
    = \lim_{r \to 0} \frac{|R_{k,\ell}(r)|^2}{|R^{(0)}_{k,\ell}(r)|^2} \,.
    \label{eq:S_def}
}
Here, $R^{(0)}_{k,\ell}$ denotes the radial wave-function without the potential term.

\begin{figure}[t]
    \centering
    \includegraphics[width=0.45\textwidth]{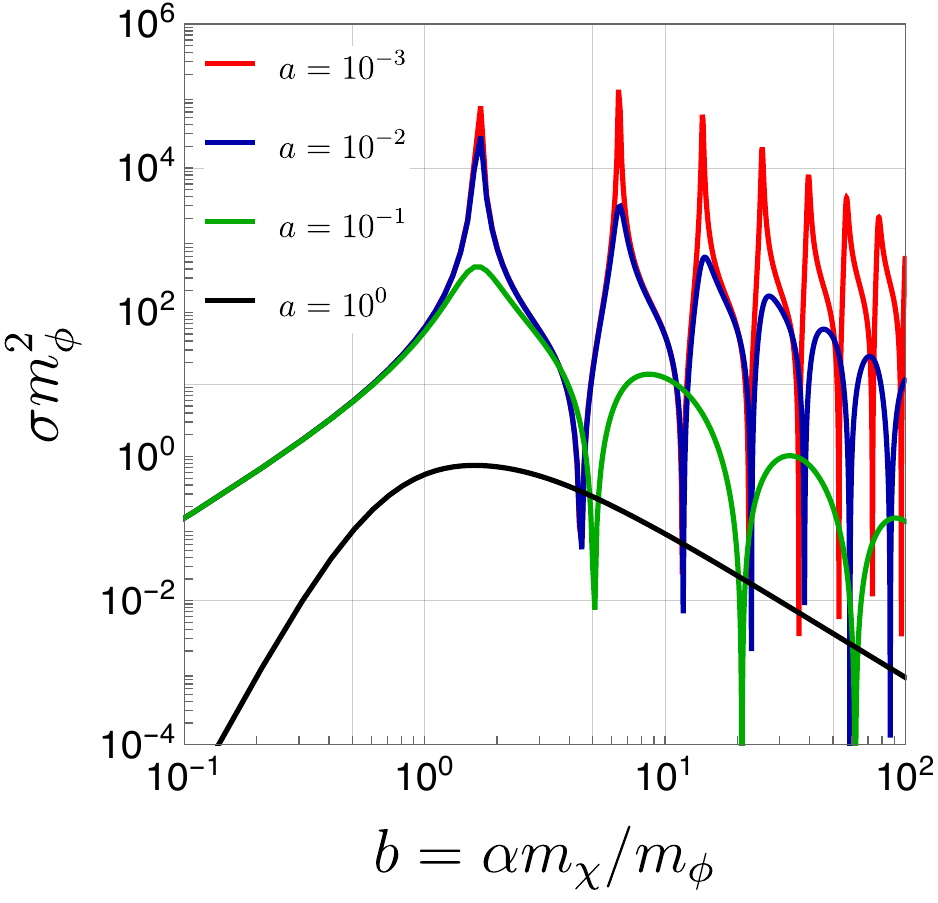}
    \includegraphics[width=0.45\textwidth]{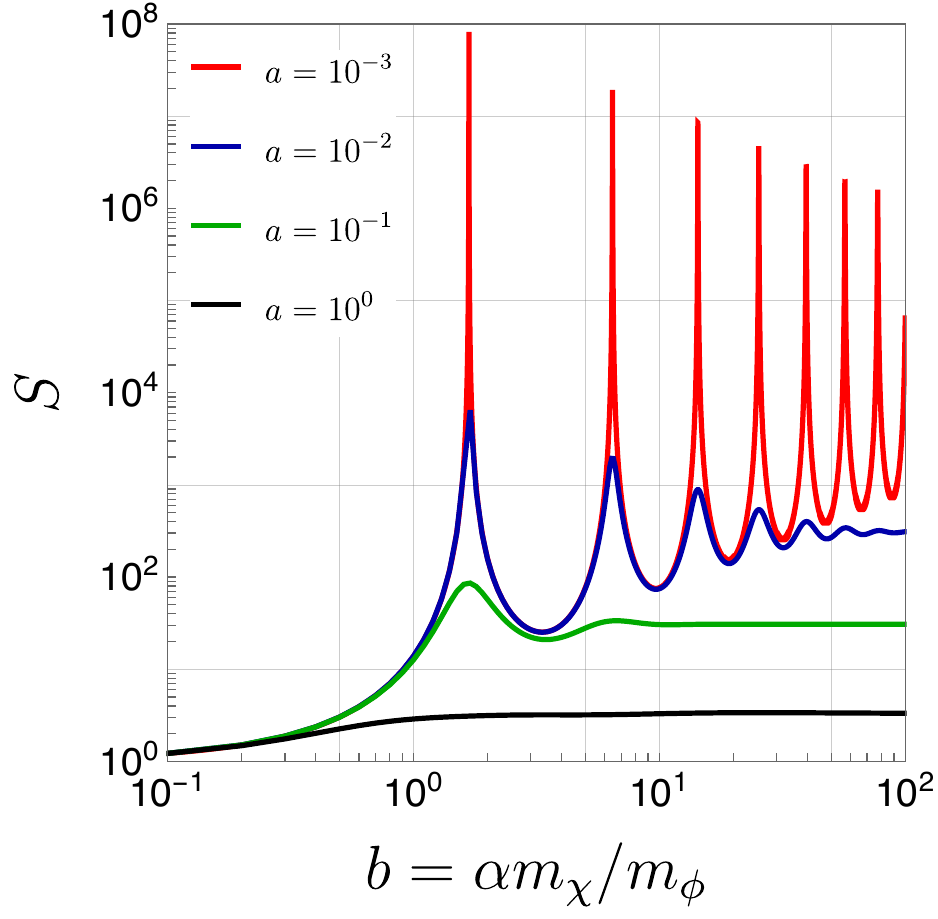}
    \caption{
        Self-scattering cross section $\sigma$ (in unit of $m_\phi$) and the Sommerfeld enhancement factor $S$ for $s$-wave processes in the Yukawa potential as a function of $b$.
    }
    \label{fig:correlation}
\end{figure}

Now, we introduce numerical methods of computing the phase shift $\delta_\ell$ and the Sommerfeld enhancement factor $S_\ell$.
Since there is no analytic solution for the Schr\"odinger equation with the Yukawa potential, we numerically solve the equation following Ref.~\cite{Tulin:2013teo} for the numerical method for obtaining the phase shift. 
We solve the one-dimensional Schr\"odinger equation \cref{eq:rad_Sch} in a certain domain $r_i \leq r \leq r_m$, covering the one where the potential term dominates the equation compared to the kinetic term and the angular-momentum term. 
Meanwhile, the asymptotic solution of the equation at large $r$ is given by 
\eqs{
    \chi_{k,\ell} (r) & \equiv r R_{k,\ell}(r) 
    \to kr e^{i \delta_\ell} \left[ \cos \delta_\ell j_\ell(kr) - \sin\delta_\ell n_\ell(kr) \right] \,,
    \label{eq:asymptoticsol}
}
where $j_\ell$ ($n_\ell$) is the spherical Bessel (Neumann) function.
Matching the numerical solution to the asymptotic solution at $r = r_m$, we obtain the phase shift $\delta_\ell$:
\eqs{
    \tan \delta_\ell 
    = \frac{k r_m j_\ell' (k r_m) - \beta_\ell j_\ell (k r_m)}{k r_m n_\ell' (k r_m) - \beta_\ell n_\ell (k r_m)} \,, \quad
    \beta_\ell 
    = \frac{r_m \chi'_{k,\ell}(r_m)}{\chi_{k,\ell}(r_m)} - 1 \,.
    \label{eq:phase_Yukawa}
}
The normalization of the wave-function is irrelevant to determine the phase shift, then we take a simple initial condition for numerically solving the Schr\"odinger equation \cref{eq:rad_Sch}.
The angular-momentum term dominates \cref{eq:rad_Sch} near the origin, and hence $\chi_{k,\ell}(r_i) \propto (k r_i)^{\ell+1}$.
Therefore, we use $\widetilde \chi_{k,\ell} \equiv A_\ell \chi_{k,\ell}$ instead of $\chi_{k,\ell}(r)$ for the numerical computation, which satisfies an initial condition $\widetilde \chi_{k,\ell}(r_i) = 1$ and $\widetilde \chi'_{k,\ell}(r_i) = (\ell+1)/r_i$.

We numerically compute the Sommerfeld enhancement factor by using the same wave-function $\widetilde \chi_{k,\ell}$.
Meanwhile, near the origin, the wave-function $\chi_{k,\ell} (r)$ asymptotically behaves as 
\eqs{
    \chi_{k,\ell} (r) \simeq B_\ell \frac{(kr)^{\ell+1}}{(2\ell+1)!!} \,.
    \label{eq:chi_at_0}
}
Here, $B_\ell$ represents the distortion of the wave-function near the origin relative to the plane-wave solution due to the presence of the potential.
In other words, the Sommerfeld enhancement factor is given by $S_\ell = |B_\ell|^2$. 
Therefore, matching the numerical solution to the asymptotic solution at $r = r_i$ (where $\widetilde\chi_{k,\ell} (r_i) = 1$), we obtain the Sommerfeld enhancement factor as
\eqs{
    S_\ell(k^2) = |B_\ell|^2 = \left| \frac{(2\ell+1)!!}{A_\ell (kr_i)^{\ell+1}} \right|^2 \,.
    \label{eq:S_Yukawa}
}
Here, we determine the normalization $A_\ell$ by using the phase shift given by \cref{eq:phase_Yukawa} and the numerical solution $\widetilde\chi_{k,\ell} (r)$ at $r = r_m$, as $A_\ell = \widetilde\chi_{k,\ell} (r_m)/\chi_{k,\ell} (r_m)$.

We show the resonant self-scattering $s$-wave cross section and Sommerfeld enhancement factor for $s$-wave annihilation process in \cref{fig:correlation}.
Both the scattering cross section and the Sommerfeld enhancement factor resonantly enhance in low velocity (\textit{i.e.} low $a$), and we can find that the resonant points for both quantities correspond with each other.
In the following sections, we clarify the underlying reason why these resonant points are closely associated with each other. 

\section{Watson's Theorem and Omn\`es Solution \label{sec:Watson}}

Now, we discuss a relation between the self-scattering phase shift and the Sommerfeld enhancement factor for the DM annihilation.
A key to demonstrate the relation is the Watson's theorem~\cite{Watson:1952ji}: the discontinuity of an amplitude is determined by the scattering phase shift of the initial (final) state in the presence of elastic scattering of the states. 

\begin{figure}[t]
    \centering
    \includegraphics[width=0.95\textwidth]{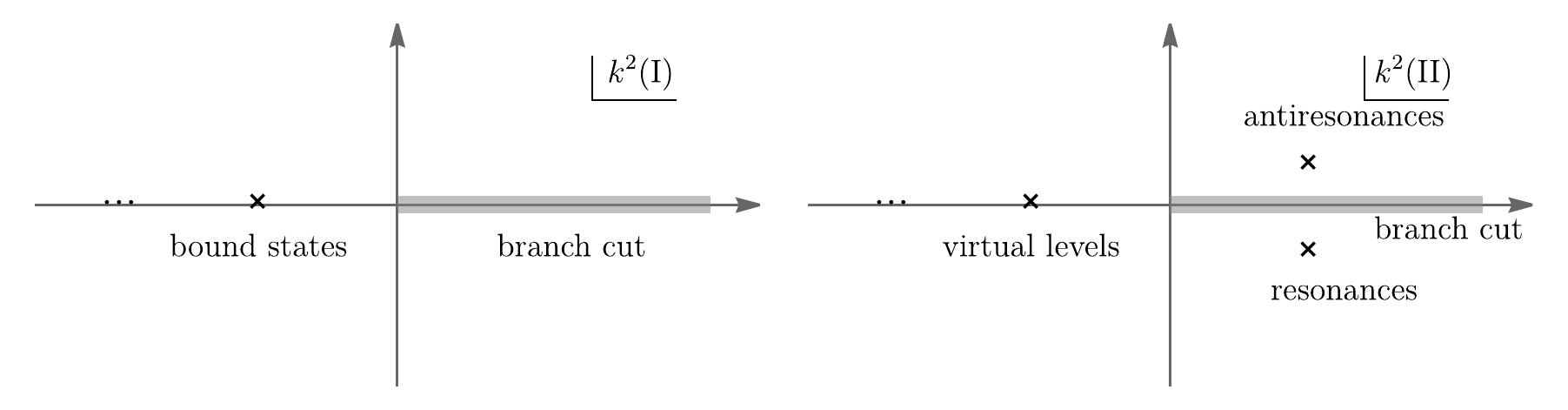}
    \caption{
    Analytic structure of the amplitude of DM annihilation in complex $k^2$ plane: (left) the first Riemann sheet ($\mathrm{Im}(k) > 0$), (right) the second Riemann sheet ($\mathrm{Im}(k) < 0$).
    }
    \label{fig:Gamma}
\end{figure}

In- and out-states are defined as eigenstates of Hamiltonian of interacting system, $H = H_0 + V$ with free Hamiltonian $H_0$ and the interaction term $V$.
These states become a free state $\Phi$ for $V \to 0$, and are formally expressed by the Lippmann-Schwinger equation
\eqs{
    | \Psi^\pm_\alpha \rangle
    = | \Phi_\alpha \rangle + \frac{1}{E - H_0 \pm i \epsilon} V | \Psi^\pm_\alpha \rangle \,,
}
where $\epsilon > 0$, and $\pm$ indicates in-state ($+$) and out-state ($-$).
$\alpha$ collectively denotes labels of the states, such as momenta, spins, and so on. 
The DM annihilation with partial-wave $\ell$ is induced via an operator $\Theta_\chi$, and we introduce the amplitude of DM annihilation as 
\eqs{
    \Gamma_\ell(k^2 + i \epsilon) \equiv \langle 0| \Theta_\chi | \Psi^+_\ell \rangle \,,
}
where $|\Psi^+_\ell \rangle$ denotes an in-state for the initial two-body system (with partial-wave $\ell$).
There is a branch cut along real $k^2$ axis for $k^2 > 0$, and there may be poles for $k^2 < 0$ corresponding to bound states in the first Riemann sheet of the $k^2$-plane as shown in \cref{fig:Gamma}.
There may also be virtual levels, resonances (unstable states), and anti-resonance in the second Riemann sheet.
The amplitude $\Gamma_\ell$ is a real function when the system has the time inversion invariance.

When the system has time-inversion invariance, by inserting the complete set of out states, we can rewrite the amplitude as follows.%
\footnote{
    Some readers may be familiar with this theorem in the following form.
    \eqs{
        \mathrm{Im} \, \Gamma_\ell(k^2+i\epsilon) = T_\ell^\ast \beta \Gamma_\ell(k^2+i\epsilon) \,.
    }
    Here, $T$-matrix element relates with the scattering phase shift as $1 + 2 i \beta T_\ell = e^{2 i \delta_\ell}$.
    $2 \beta$ denotes the relative velocity of the two-body state. 
}
\eqs{
    \Gamma_\ell(k^2 + i \epsilon)
    =\langle 0| \Theta_\chi | \Psi^-_\ell \rangle \langle \Psi^-_\ell | \Psi^+_\ell \rangle 
    = e^{2 i \delta_\ell} \Gamma_\ell(k^2 - i \epsilon)
    = e^{2 i \delta_\ell} \Gamma_\ell(k^2 + i \epsilon)^\ast \,.
    \label{eq:Watson}
}
Here, $e^{2 i \delta_\ell}$ originates from the $S$-matrix element:
\eqs{
    e^{2 i \delta_\ell} = \langle \Psi^-_\ell | \Psi^+_\ell \rangle\,,
}
and $\delta_\ell$ is the phase shift of the $\chi\chi \to \chi\chi$ scattering process.
This indicates that the discontinuity of the amplitude is given by the scattering phase $\delta_\ell$, which is known as the Watson's theorem~\cite{Watson:1952ji}.
The analytic solution of \cref{eq:Watson} is known as the Omn\`es solution~\cite{muschelivsvili1953singular,Omnes:1958hv}.
The form of the solution is given by
\eqs{
    \Gamma_\ell (k^2 + i \epsilon) & = F_\ell(k^2) e^{\omega_\ell(k^2) + i \delta_\ell(k)} \,, \\
    F_\ell(k^2) & = \prod_{b} \frac{k^2}{k^2+\kappa_b^2} \,, \quad 
    \omega_\ell(k^2) = \frac{1}{\pi} \mathop{\,\text{--}\hspace{-3.4mm}\int}\limits_0^\infty \frac{\delta_\ell(p)}{p^2 - k^2} d p^2 \,.
    \label{eq:OmnesF}
}
Here, $F_\ell(k^2)$ is a real rational function, which is responsible for the pole structure of $\Gamma_\ell(k^2)$ in the first Riemann sheet in \cref{fig:Gamma}, namely the bound states.%
\footnote{
    We may also have resonances, but these contributions are somehow automatically incorporated in the Omn\`es function as we will see later. 
}
$b$ denotes the label of bound states.
In the presence of bound states in the $\chi\chi \to \chi\chi$ scattering, $F_\ell(k^2)$ has poles along the positive imaginary axis on the complex $k$ plane. 
For instance, $F_\ell(k^2)$ is proportional to $(k^2+\kappa_b^2)^{-1}$ when there is a pole at $k = i \kappa_b$.
We will discuss the numerator after introducing the Levinson's theorem. 
The factor $e^{\omega_\ell(k^2)+ i \delta_\ell(k)}$, called as the Omn\`es function, reproduces the branch cut of the amplitude.%
\footnote{
    We note that the function $\omega_\ell(k^2)+ i \delta_\ell(k)$ is analytically continued to complex $k^2$ as $\omega_\ell(k^2)$ in \cref{eq:OmnesF} but replacing the principal value integral by a normal integral.
} 

The Levinson's theorem~\cite{Levinson:407337} relates the behavior of the scattering phase shift to the number of bound states.
We will introduce an intuitive proof of the theorem in \cref{app:Levinson}.
The theorem states that
\eqs{
    \delta_\ell(0) - \delta_\ell(\infty) = \left(n_\ell + \frac12 N \right) \pi \,,
    \label{eq:Levinson}
}
where $n_\ell$ denotes the number of the bound states, and $N = 1$ only for $s$-wave scattering just on the zero-energy resonance and $N = 0$ for other situations (other partial-waves or off the zero-energy resonance).
As we will see in detail in the next section, there are zero-energy states on the resonances, which we refer to the zero-energy resonances: they are non-normalizable (\textit{i.e.,} virtual levels) for $s$-wave%
\footnote{
    For Hulth\'en potential, we can find that the states corresponding to $s$-wave zero-energy resonances can be virtual levels since we have analytical solution for the Schr\"odinger equation (see \cref{app:Hulthen}).
}%
, while they are normalizable (\textit{i.e.,} bound states) for $\ell \geq 1$.
$n_\ell$ counts only bound states, but not virtual levels.
$\delta_\ell(0)/\pi$ is closely related to the number of bound states due to our normalization of $\delta_\ell$, $\delta_\ell(\infty) = 0$. 

We determine the numerator of $F_\ell(k^2)$ as follows.
We choose the normalization of the amplitude ignoring the potential term to be $\Gamma^{(0)}_\ell (k^2) = 1$.
The amplitude $\Gamma_\ell (k^2)$ corresponds with $\Gamma^{(0)}_\ell (k^2)$ at high velocity. 
This determines the behavior of $F_\ell(k^2)$ in the limit of $k^2 \to \infty$.
$\omega_\ell(k^2)$ goes to zero in this limit since we take the normalization $\delta_\ell(\infty) = 0$, and hence the function $F_\ell(k^2)$ satisfies $F_\ell(k^2 \to \infty) = 1$.
Meanwhile, according to the Levinson's theorem~\cite{Levinson:407337}, the phase shift at small $k$ relates with the number of the bound states.
The $\omega_\ell (k^2)$ logarithmically diverges at small $k$, $\omega_\ell (k^2) \simeq \pi^{-1} \delta_\ell(0) \ln (\Lambda^2/k^2)$, since the integral (\ref{eq:OmnesF}) is dominated at $p^2 = k^2$. 
$\Lambda$ characterizes the maximum momentum up to which the phase shift is regarded to be constant.
Hence, the Omn\`es function behaves as 
\eqs{
    e^{\omega_\ell(k^2)} \simeq \left( \frac{\Lambda^2}{k^2} \right)^{\delta_\ell(0)/\pi} \simeq \left( \frac{\Lambda^2}{k^2} \right)^{n_\ell} \,,
    \label{eq:OmnesFlowk}
}
at small $k$ with the number of bound states $n_\ell$.
Since the Sommerfeld enhancement factor should be constant at small $k$ except for the zero-energy resonance (we will discuss later), the functional form of $F_\ell(k^2)$ satisfies $F_\ell(k^2) \sim (k^2)^{n_\ell}$ at small $k$.%
\footnote{
    The analyticity of the scattering amplitude implies the low-velocity scaling as $f_\ell(k^2) \sim k^{2 \ell}$ and the annihilation cross section should scale as $\sigma_{\mathrm{ann},\ell} v \sim v^{2 \ell}$.
    This low-$k$ dependence of the annihilation cross section is already included in the annihilation cross section without potential $(\sigma_{\mathrm{ann},\ell} v)_0$. 
}
Hence, we obtain the functional form of $F_\ell(k^2)$ as in \cref{eq:OmnesF}.
Combining the behavior of the Omn\`es function at small $k$, we have a constant $\Gamma_\ell(k^2)$ at small $k$ except for the zero-energy resonance ($\kappa_b = 0$).
For our formulation, we need to know the number of the bound states (\textit{i.e.}, the low-energy behavior of the phase shift) and the position of the bound states $\kappa_b$. 
We discuss them in detail by introducing the effective range theory in the next section.

Finally, the Sommerfeld enhancement factor is obtained as a ratio of the velocity-weighted annihilation cross sections with and without potential, and hence it may be rewritten as the ratio of the squared amplitudes with and without potential.
\eqs{
    S_\ell(k^2) 
    = \frac{|\Gamma_\ell (k^2)|^2}{|\Gamma_\ell^{(0)} (k^2)|^2} 
    = |\Gamma_\ell (k^2)|^2\,.
    \label{eq:WatsonS}
}
We may compute $S_\ell$ for a given phase shift by the use of this formula instead of the standard computational method of the Sommerfeld enhancement factor given by \cref{eq:S_def}.
In \cref{sec:Yukawa}, we compare the factors computed in these two ways.

\section{Phase Shift and Bound States \label{sec:ERT}}

Our formulation requires to know the number of bound states and corresponding poles $\kappa_b$ in order to determine $F_\ell(k^2)$ as introduced in the previous section. 
The number of bound states is closely related to the low-energy scattering phase shift via the Levinson's theorem. 
It is a key for our formulation to know the low-energy behavior of the phase shift and the poles corresponding to the bound states.
The low-energy scattering of two-body system is approximately described by the effective range theory~\cite{Bethe:1949yr,Blatt:1949zz}, in which only two model parameters describe the two-body scattering. 
We determine the poles of the annihilation amplitudes by the use of the effective range theory as far as the poles corresponds to the shallow bound states, as we will see later.
We also use the effective range theory in order to find the reference points for our analysis. 

It is a key point for the effective range theory that the wave-function of the two-body system must be analytic at $k = 0$. 
This analytic property determines the low-energy behavior of the phase shift, and hence the phase shift $\delta_\ell$ for multipole $\ell$ is expanded in the low-energy as 
\eqs{
    k^{2\ell+1} \cot \delta_\ell(k) \simeq - \frac{1}{a_\ell^{2\ell+1}} + \frac{1}{2 r_{e,\ell}^{2 \ell-1}} k^2 \,.
    \label{eq:ERT}
}
Here, $a_\ell$ and $r_{e,\ell}$ are called the scattering length and the effective range, respectively.
This expansion is valid as far as $|k| < |r_{e,\ell}|^{-1}$, otherwise the $k^4$ and higher terms are not negligible and it is beyond the effective range theory. 
This formalism has been recently utilized as a model-independent approach in order to approximate the velocity-dependent self-scattering of DM in the context of self-interacting DM~\cite{Chu:2019awd} (see Refs.~\cite{Kang:2020afi,Kamada:2020buc,Cline:2022leq,Kondo:2022lgg} for concrete models).

\begin{figure}[t]
    \centering
    \includegraphics[width=0.45\textwidth]{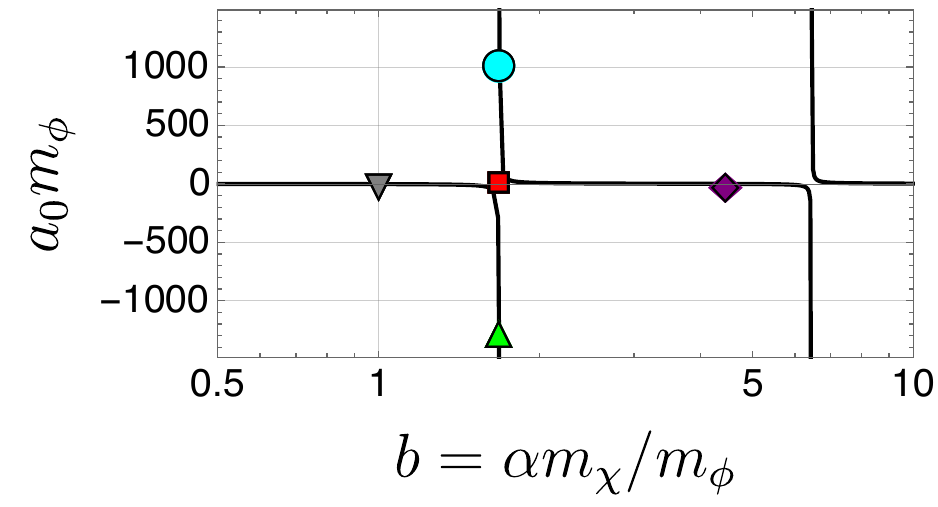}
    \includegraphics[width=0.45\textwidth]{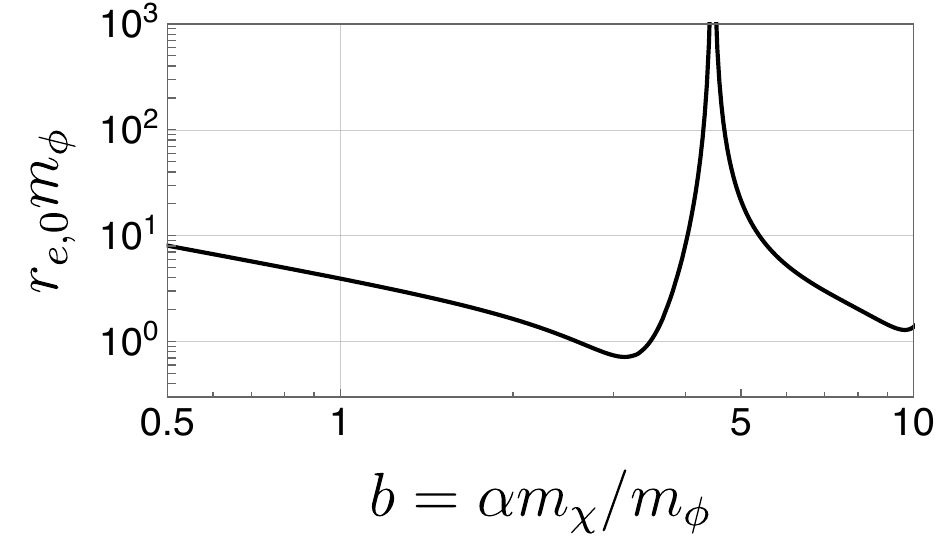}
    \includegraphics[width=0.45\textwidth]{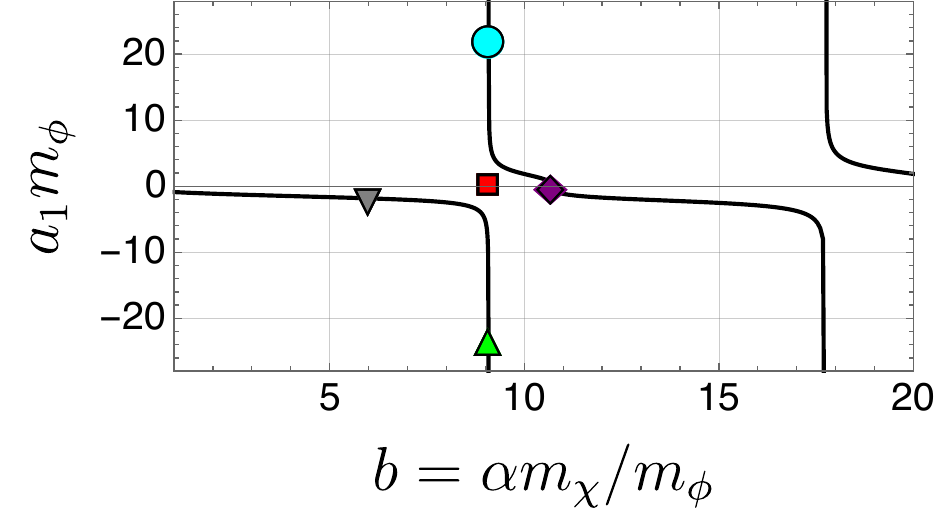}
    \includegraphics[width=0.45\textwidth]{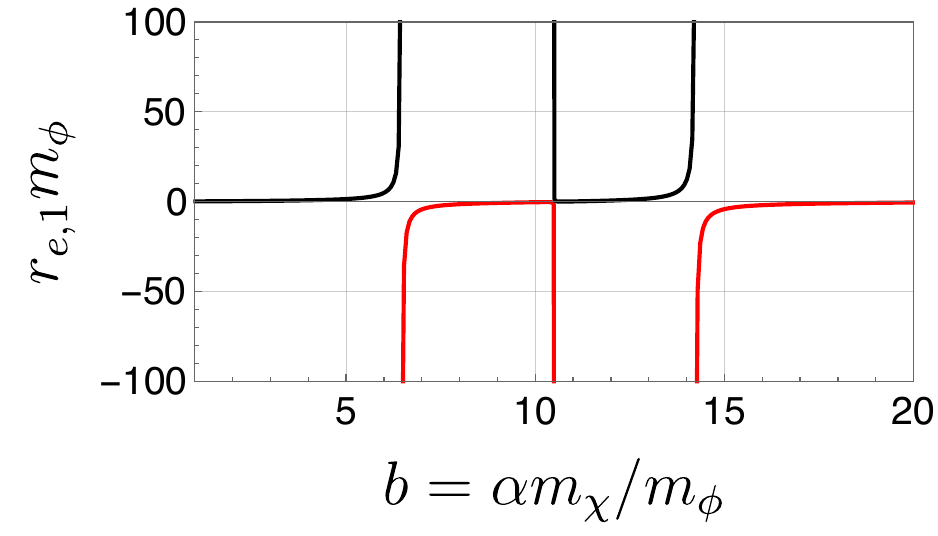}
    \caption{
        (\textit{top}): Scattering length $a_0$ and effective range $r_{e,0}$ for $s$-wave scattering in the Yukawa potential as a function of $b$ in unit of $\delta$.
        Since the scattering length diverges at $b = 1.680$, we put the mark for the point on the $b$-axis. 
        (\textit{bottom}): Same as top panels, but for $p$-wave processes. 
        We put the mark for the point $b = 9.082$ on the $b$-axis. 
    }
    \label{fig:Yukawa_sl}
\end{figure}

Using the effective range theory, we find poles of the scattering amplitude given by \cref{eq:sc_amp_decom}. 
The poles of the $s$-wave amplitude are found along the imaginary axis of the complex $k$-plane as far as $2 r_{e,0} < a_0$.
One of the poles that is closer to the real axis, denoted by $k = i \kappa_b$, is given by
\eqs{
    \kappa_b = \frac{1}{r_{e,0}} \left( 1 - \sqrt{1 - \frac{2 r_{e,0}}{a_0}} \right) \,,
}
and $\kappa_b \simeq a_0^{-1}$ for $|a_0| \gg |r_{e,0}|$.
We again note that this expression is valid as far as $|\kappa_b| < |r_{e,0}|^{-1}$.
The corresponding energy $E_\mathrm{pole} = - \kappa_b^2/2\mu$ is negative, and hence there exists a bound state ($a_0 > 0$) or a virtual level ($a_0 < 0$; a bound state with non-normalizable wave-function).
These states become a zero-energy state as $|a_0| \to \infty$ ($\kappa_b \to 0$), and we refer to these points such that $|a_\ell| \to \infty$ as the zero-energy resonances.
For the higher partial-waves ($\ell \geq 1$), the poles appear at $\kappa_b^2 \simeq -2 r_{e,\ell}^{2\ell-1}/a_\ell^{2\ell+1}$, and hence there exist both a bound state and a virtual level (for $a_\ell > 0$ and $r_{e,\ell} < 0$, or for $a_\ell < 0$ and $r_{e,\ell} > 0$).
The phase shift has interesting behavior for the effective-range theory parameters in $a_\ell < 0$ and $r_{e,\ell} < 0$.
For such parameter range, there is a quick change of the phase shift by $\pi$ at a certain momentum, $k^2 = 2 r_{e,\ell}^{2\ell-1}/a_{\ell}^{2\ell+1}$, and the phase shift goes through $\pi/2$ (or a half-integer multiple of $\pi$).
This indicates that the cross section has a peak at the momentum, namely there exists a Breit-Wigner type resonance.

We show the scattering length and the effective range in \cref{fig:Yukawa_sl} for $s$-wave (top panels) and for $p$-wave (bottom panels).
We mark several reference points in the left panels, which we will focus on in \cref{sec:Yukawa}. 
These reference points are classified into three types: points below the first zero-energy resonance ($\blacktriangledown$ and $\blacktriangle$), on the zero-energy resonance ($\blacksquare$), and points above the zero-energy resonance (\tikzcircle{4pt} and $\blacklozenge$). 
Since the scattering length $a_\ell$ diverges on the zero-energy resonance, we put marks $\blacksquare$ ($b = 1.680$ for $s$-wave and $b = 9.082$ for $p$-wave) on $b$-axis in the figure.
Meanwhile, the scattering length vanishes, $a_\ell = 0$, at certain points depicted as $\blacklozenge$ in the figure ($b = 4.466$ for $s$-wave and $b = 10.6997$ for $p$-wave), and we call the points as the (first) anti-resonance.
The effective range theory is no longer valid at these points since $1/a_\ell$ is indefinite and $1/|r_{e,\ell}| \to 0$. 
For both partial-waves, $a_\ell$ is negative for $b$ below the zero-energy resonance, while $a_\ell$ is positive for $b$ above the zero-energy resonance up to the anti-resonance.
The effective range $r_{e,\ell}$ is always positive for $s$-wave, while $r_{e,\ell}$ takes both positive and negative values for $p$-wave. 
We use different colors for depicting connected lines for $r_{e,1}$ in the bottom-right panel.
On the point $\blacktriangle$ for $p$-wave, since both $a_1$ and $r_{e,1}$ are negative, there exists a Breit-Wigner type resonance. 

\begin{figure}[t]
    \centering
    \includegraphics[width=0.45\textwidth]{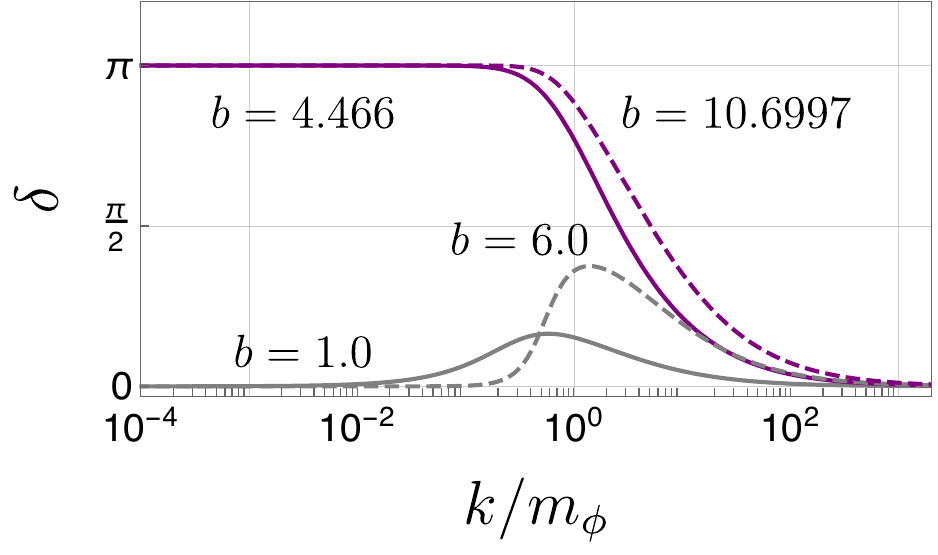}
    \includegraphics[width=0.45\textwidth]{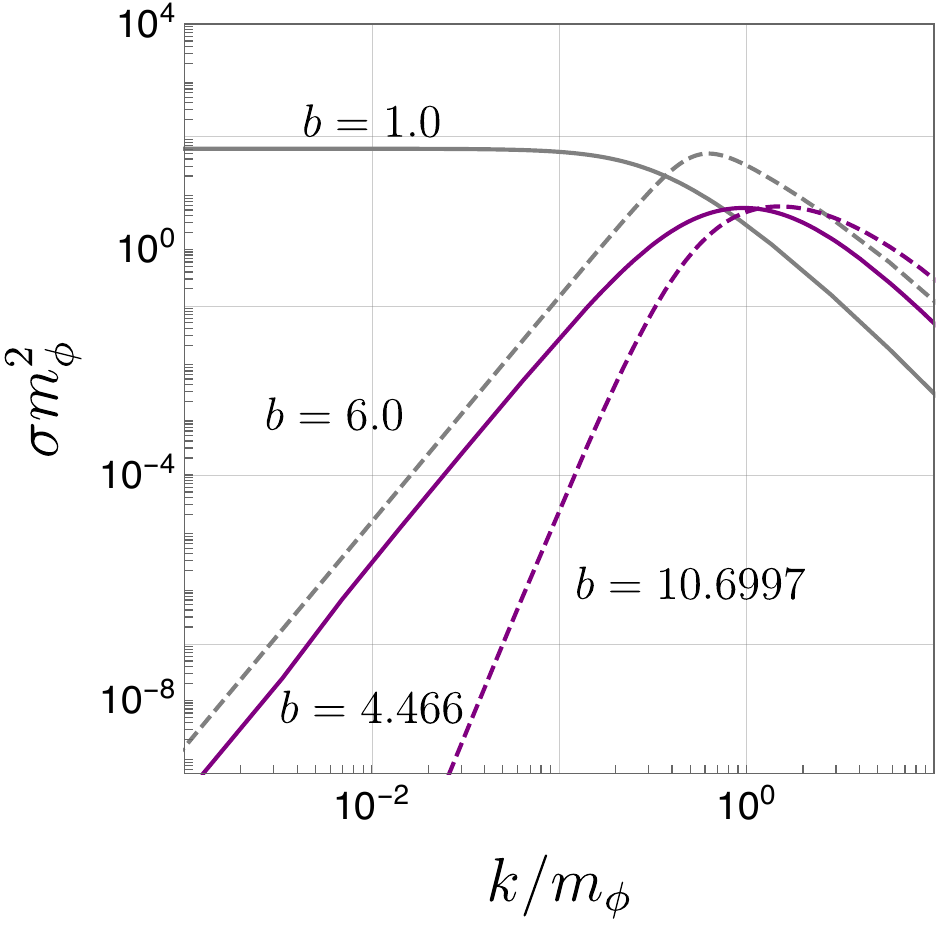}
    \caption{
        Phase shift and cross section under the Yukawa potential for $s$-wave (solid) and $p$-wave (dashed) processes.        
        (\textit{left}): 
        Phase shift for given $b$ as a function of $k/m_\phi$. 
        (\textit{right}): 
        Cross section in unit of $m_\phi$ for given $b$ as a function of $k/m_\phi$. 
    }
    \label{fig:Yukawa}
\end{figure}

Now, we discuss the phase shift and the cross section for $s$- and $p$-wave scattering at points marked as $\blacktriangledown$: $b = 1.0$ for $s$-wave and $b = 6.0$ for $p$-wave.
For $s$-wave scattering, the scattering length is negative ($a_0 < 0$) and the effective range is positive ($r_{e,0} > 0$).
The effective range theory \cref{eq:ERT} tells us that $\delta_0 \to + 0$ in two limits $k \to 0$ and large $k$ and that $\delta_0$ is maximized at $k \simeq (- a_0 r_{e,0})^{-1/2}$.
We show the behavior of the phase shift and the cross section in \cref{fig:Yukawa} as gray-solid lines.
We also show the scattering cross section (in unit of $m_\phi^2$) in the right panel.
The cross section approaches to a constant value $\sigma \sim 4 \pi a_0^2$ in small $k$ limit.
Meanwhile, for $p$-wave scattering, the scattering length is negative while the effective range is positive, and $\delta_1$ is maximized at $k \simeq (- r_{e,1}/a_1^3)^{1/2}$.
The gray-dashed lines in \cref{fig:Yukawa} show the behavior of the phase shift and the cross section for $b = 6.0$ of $p$-wave scattering.
The phase shift scales as $\cot \delta_1 \simeq (- a_1 k)^{-3}$ at small $k$, and hence the cross section scales as $k^4$.

\cref{fig:Yukawa} also shows the phase shift and the cross section for the first anti-resonance points marked as $\blacklozenge$: $b = 4.466$ for $s$-wave and $b = 10.6997$ for $p$-wave.
We note again that the effective range theory is no longer valid at these points.
For $s$-wave scattering, the phase shift approaches to $\pi$ at small $k$, shown as a purple-solid line in the left panel of \cref{fig:Yukawa}.
More precisely, in the small $k$ limit, we find that the phase shift behaves as $\delta_0 \simeq \pi - b_0^3 k^3$ with $b_0$ being a length parameter, and hence the cross section scales as $k^4$ at small $k$.
Meanwhile, $p$-wave scattering, we find that the phase shift behaves as $\delta_1 \simeq \pi - b_1^5 k^5$ with $b_1$ being a length parameter, shown as a purple-dashed line in the left panel of \cref{fig:Yukawa}.
The cross section scales as $k^8$ from the $k$-dependence of the phase shift at small $k$.
As a generalization, for the first anti-resonance of multipole $\ell$, we expect that the phase shift behaves as $\delta_\ell \simeq \pi - (b_\ell k)^{2 \ell+3}$ at small $k$ with $b_\ell$ being a length parameter.
Therefore, the effective range theory does not work since $k^{2 \ell+1} \cot \delta_\ell \sim b_\ell^{-2\ell-3} k^{-2}$, and the scattering cross section scales as $k^{4 \ell + 4}$ at small $k$.

\begin{figure}[t]
    \centering
    \includegraphics[width=0.45\textwidth]{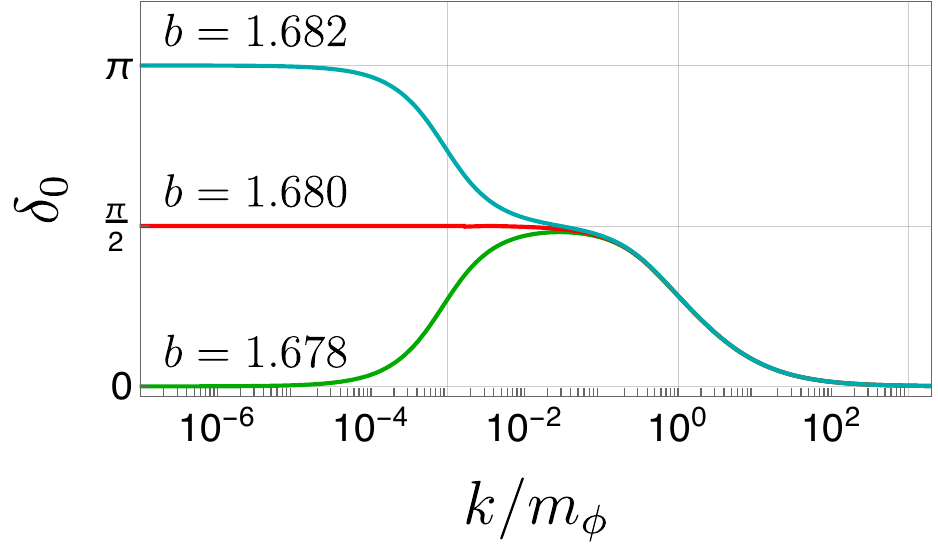}
    \includegraphics[width=0.45\textwidth]{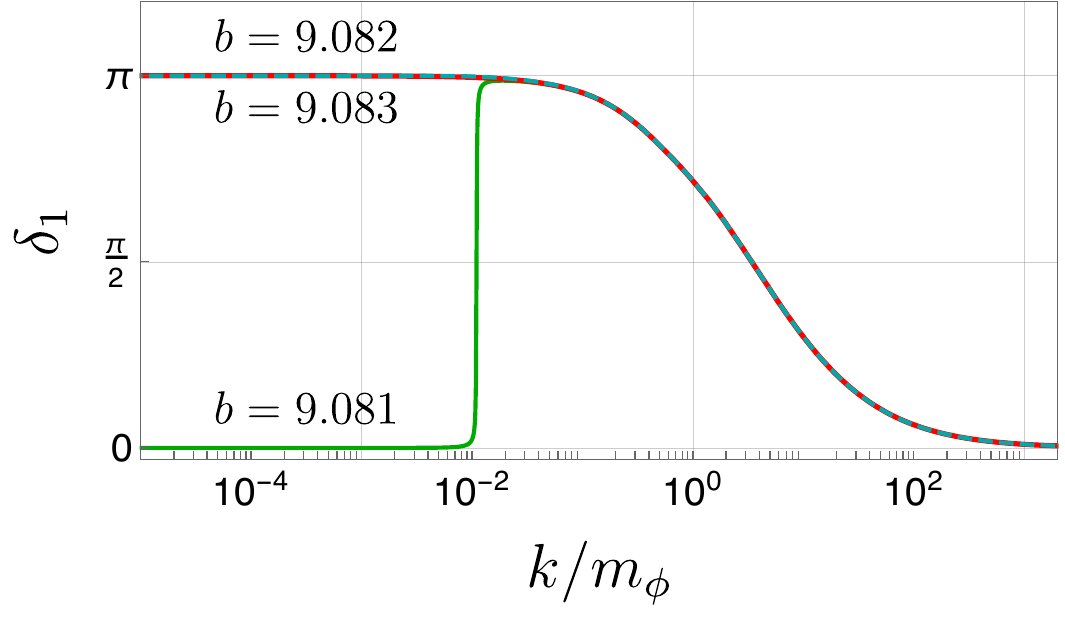}
    \includegraphics[width=0.45\textwidth]{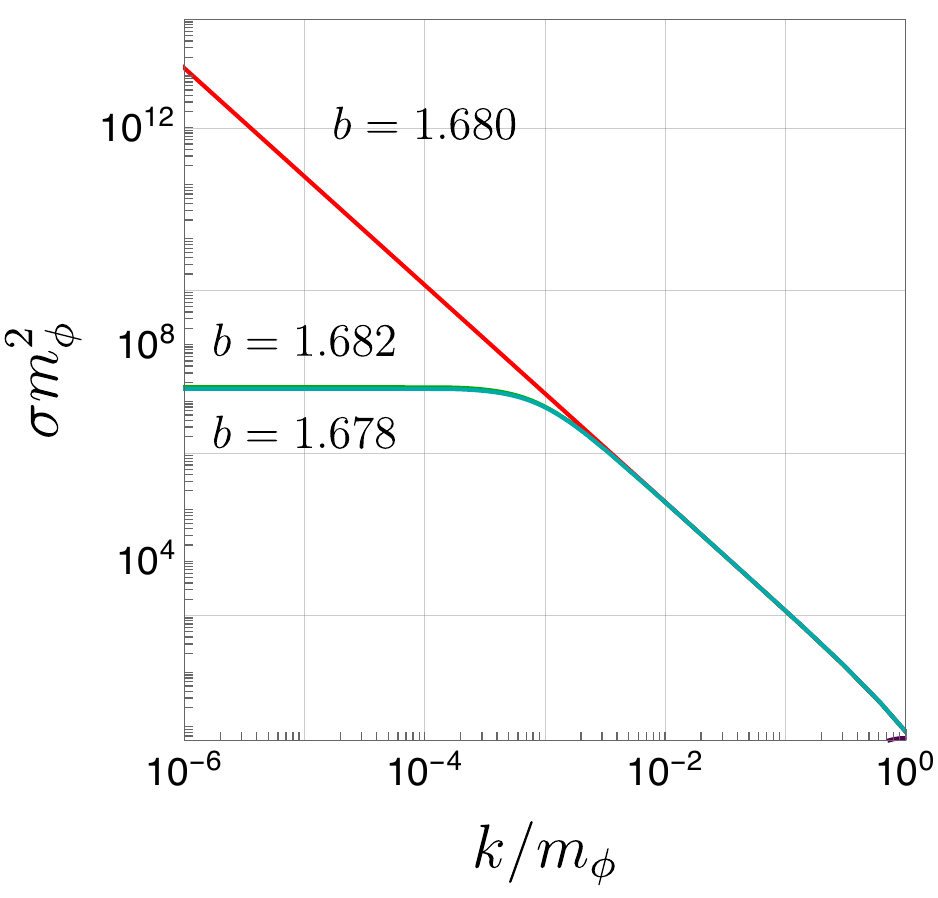}
    \includegraphics[width=0.45\textwidth]{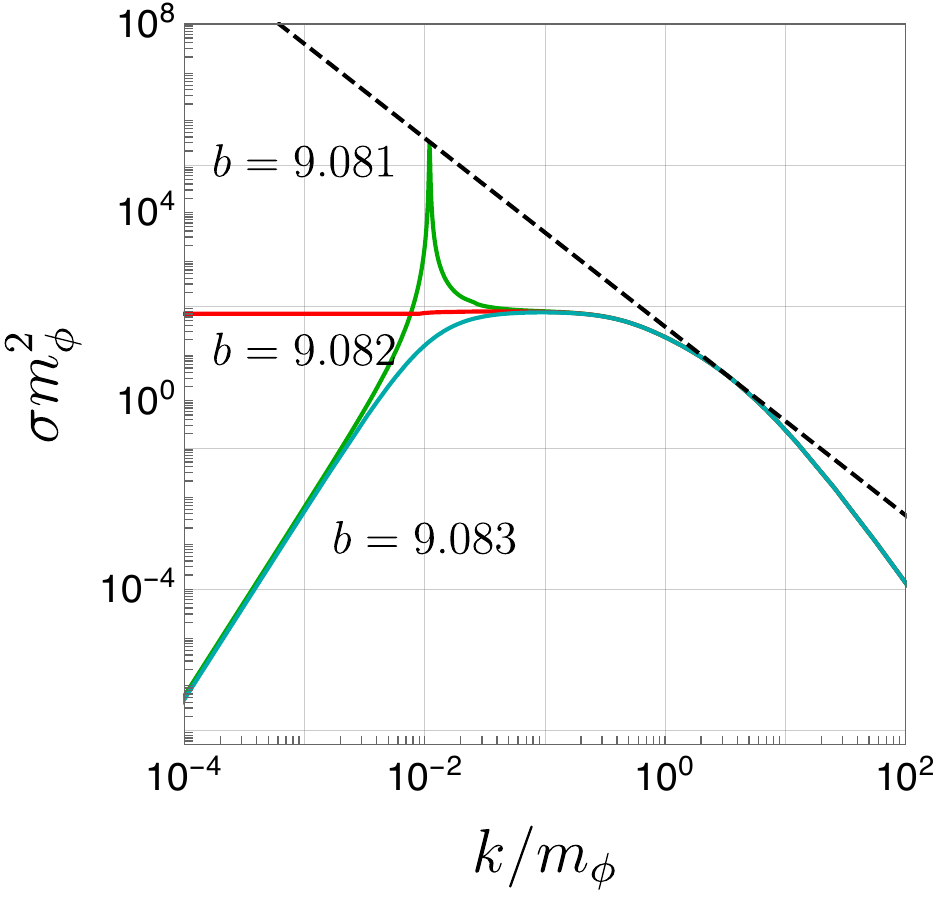}
    \caption{
        Phase shift and cross section near the first zero-energy resonance.
        (\textit{top}): Phase shifts of $s$-wave scattering (left) and $p$-wave scattering (right) for given $b$ as a function of $k/m_\phi$. 
        (\textit{bottom}): 
        Cross section of $s$-wave scattering (left) and $p$-wave scattering (right) in unit of $m_\phi$ for given $b$ as a function of $k/m_\phi$. 
        The black-dashed line in the bottom-right panel shows the cross section $\sigma \simeq 3 \times 4 \pi/k^2$.
    }
    \label{fig:Yukawa2}
\end{figure}

In \cref{fig:Yukawa2}, we show the phase shift and the scattering cross section for $s$-wave (top panels) and $p$-wave (bottom panels) near the first zero-energy resonance.
Below the zero-energy resonance, the phase shift (depicted as green lines) goes to zero in two limits $k \to 0$ and large $k$, and hence the phase shift takes a maximum value at a certain $k$.
Therefore, the cross section (depicted as green lines) approaches to a constant value $\sigma \sim 4 \pi a_0^2$ at small $k$ for the $s$-wave scattering, and the cross section scales as $k^4$ at small $k$ for the $p$-wave scattering.
The $p$-wave scattering cross section has a spiky behavior at $k \sim 10^{-2} m_\phi$ for $b = 9.081$.
Since both scattering length and effective range are negative for this $b$, the phase shift takes $\pi/2$ at $k = (2 r_{e,1}/a_1^3)^{1/2}$ according to the effective range theory. 
Therefore, the scattering cross section takes its maximum value $\sigma \simeq 3 \times 4 \pi/k^2$ at this point.
This behavior is also understood by the use of the resonant scattering cross section with running decay width~\cite{Chu:2018fzy,Chu:2019awd}, 
\eqs{
    \sigma_\ell 
    = \frac{4 \pi (2 \ell +1)}{k^2} \frac{\Gamma^2(E)/4}{(E-E_R)^2 + \Gamma^2(E)/4}\,, 
    \label{eq:crosssection_R}
}
with $\Gamma(E) = m_R \gamma v^{2 \ell +1}$ and $m_\chi E_R = k_R^2 = 2 r_{e,\ell}^{2 \ell -1} /a_\ell^{2 \ell + 1}$.
We find a relation between the effective-range theory parameters and the running decay width by matching the cross sections \cref{eq:crosssection} and \cref{eq:crosssection_R} near the resonance $E \simeq E_R$ as follows:
\eqs{
    a_\ell^{2 \ell + 1} = - \frac{\Gamma(E_R)}{2 E_R k_R^{2 \ell + 1}} \,, \qquad
    r_{e,\ell}^{2 \ell - 1} = - \frac{m_\chi \Gamma(E_R)}{4 k_R^{2 \ell + 1}} \,.
}
The cross section \cref{eq:crosssection_R} behaves as $\sigma_\ell \simeq 4 \pi (2 \ell + 1) a_\ell^{4 \ell + 2} k^{4 \ell}$ at low energy, which agrees with the low-energy behavior of the cross section \cref{eq:crosssection}.
We confirm that the cross section \cref{eq:crosssection_R} is in agreement with the green line in the right-bottom panel of \cref{fig:Yukawa2} as far as $k \lesssim |r_{e,1}|^{-1}$.

The phase shift goes to $\pi$ at small $k$ for $b$ above the first zero-energy resonance for $s$- and $p$-waves, which is depicted as cyan lines in \cref{fig:Yukawa2}.
More precisely, the phase shift behaves as $\delta_0 \simeq \pi - a_0 k$ for $s$-wave, and $\delta_1 \simeq \pi - a_1^3 k^3$ for $p$-wave. 
There is a bound state; this is because the scattering length is positive at this point for $s$-wave, and $a_1 > 0$ and $r_{e,1} < 0$ for $p$-wave.
This also confirms the Levinson's theorem (with $n_\ell = 1$), as we will introduce in the next section. 
The scattering cross section approaches to the constant value, $\sigma \simeq 4 \pi a_0^2$, for $s$-wave, and the scattering cross section scales as $k^4$ for $p$-wave.

Meanwhile, the phase shift approaches to $\pi/2$ for $s$-wave and $\pi$ for $p$-wave on the first zero-energy resonance (depicted as red lines in \cref{fig:Yukawa2}).
For these points, the low-energy behavior of the phase shift is controlled only by the effective range since the scattering length diverges. 
Namely, at small $k$, the phase shift behaves as $2 \delta_0 \simeq \pi - r_{e,0} k$ for $s$-wave and as $\delta_1 \simeq \pi - 2 r_{e,1} k$ for $p$-wave.
The scattering cross section scales as $k^{-2}$ for $s$-wave, and the cross section approaches to a constant value for $p$-wave.

\section{Comparison of Sommerfeld Enhancement Factors \label{sec:Yukawa}}

Let us now compare the Sommerfeld enhancement factors computed in two ways: directly computed via \cref{eq:S_Yukawa} and computed using the Omn\`es solution [\cref{eq:WatsonS}] for a given phase shift \cref{eq:phase_Yukawa}.
We consider $s$- and $p$-waves here, but we also give a discussion of the first zero-energy resonance for $d$-wave in \cref{app:dwave}.

\begin{figure}[t]
    \centering
    \includegraphics[width=0.45\textwidth]{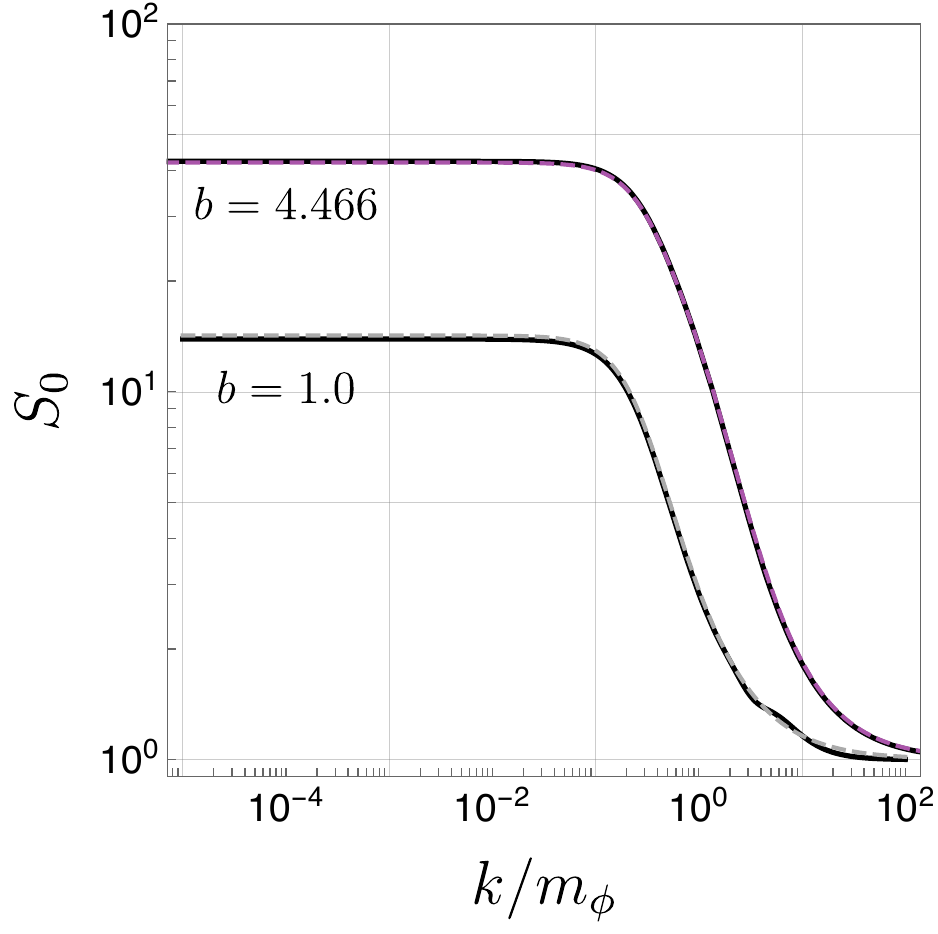}
    \includegraphics[width=0.45\textwidth]{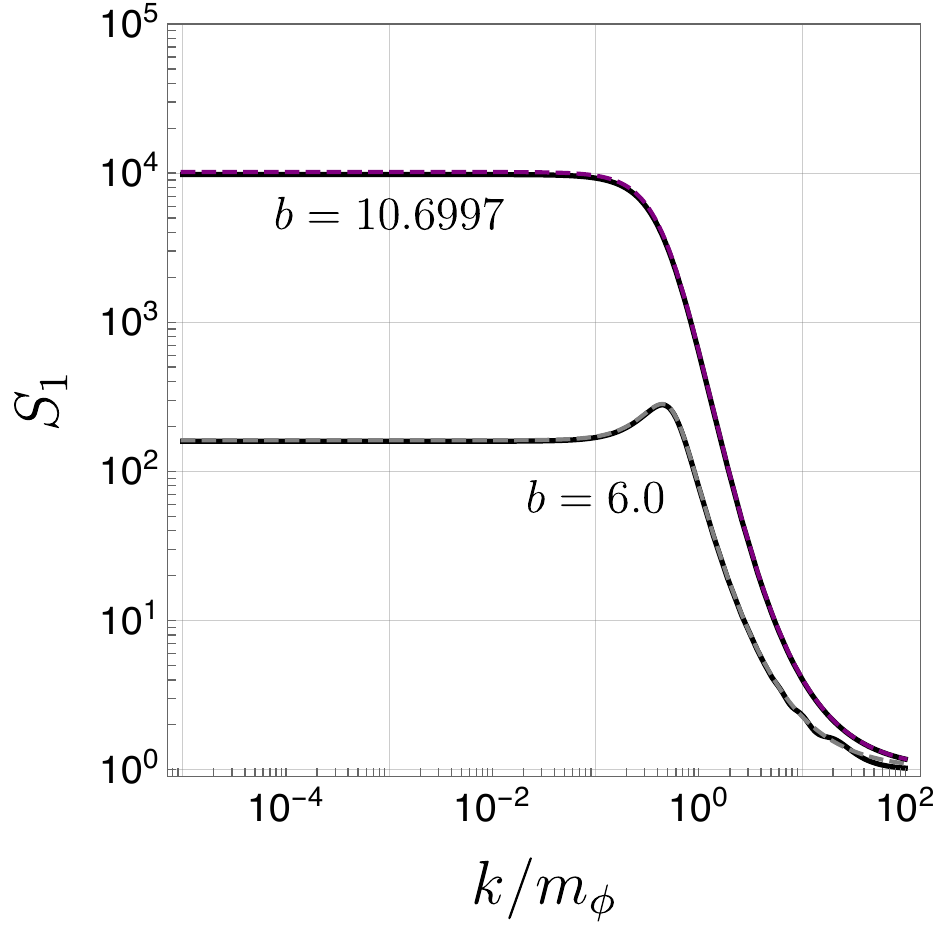}
    \caption{
        Sommerfeld enhancement factors for given $b$ as a function of $k/m_\phi$: (\textit{left}) for $s$-wave, while (\textit{right}) for $p$-wave.
        $S$ from direct computation is shown as black-solid lines, while $S$ from the phase shift and the Omn\`es function is shown as colored-dashed lines.
    }
    \label{fig:YukawaS}
\end{figure}

First, we discuss the reference points far from the first zero-energy resonance.
Concerning the points marked as $\blacktriangledown$ in \cref{fig:Yukawa_sl}, the function $F_\ell(k^2) = 1$ due to the absence of bound states.
Since $p \in [(-1/a_\ell)\,,1/r_{e,\ell}]$ dominates the integral given by \cref{eq:OmnesF}, $\omega_\ell(k^2)$ goes to a non-zero finite value at small $k$, and the Omn\`es solution $\Gamma_\ell$ also goes to a finite value in the limit. 
As for the anti-resonance points marked as $\blacklozenge$ in \cref{fig:Yukawa_sl}, the phase shift approaches to $\pi$ at small $k$.
The small-$k$ behavior of $\omega_\ell (k^2)$ is determined by the phase shift at low energy, and the Omn\`es function behaves as,
\eqs{
    e^{\omega_\ell(k^2)}
    \simeq \left( \frac{\Lambda^2}{k^2} \right)^{\delta_\ell(0)/\pi} = \frac{\Lambda^2}{k^2} \,,
    \label{eq:smallk1}
}
in the small $k^2$ limit with a mass-dimension one parameter $\Lambda$ as shown in \cref{eq:OmnesFlowk}.
$\Lambda$ characterizes the maximum scale where the phase shift is considered to be constant.
$\Lambda \simeq b_{\ell}^{-1}$ since we focus on the anti-resonance points here.%
\footnote{
    From the low-$k$ behavior of the phase shift, discussed in the previous section, we expect that $\Lambda \simeq a_\ell^{-1}$ for other off-resonance points, and $\Lambda \simeq r_{e,\ell}^{-1}$ for the zero-energy resonances.
}  
Meanwhile, since there is a bound state for the parameter, $F_\ell(k^2)$ is proportional to $k^2$.
Hence, the Omn\`es solution $\Gamma_\ell$ goes to a finite value in the limit as required by the partial-wave unitarity.

\cref{fig:YukawaS} shows the Sommerfeld enhancement factors computed in two different ways: $S_\ell$ computed directly via \cref{eq:S_Yukawa} shown as black-solid lines, and $S_\ell$ computed with the Omn\`es function shown as color-dashed lines. 
We choose the position of a bound state to be $\kappa_b = 1.32 m_\phi$ for $b = 4.466$ ($s$-wave) and $\kappa_b = 0.66 m_\phi$ for $b = 10.6997$ ($p$-wave).
We find that two $S_\ell$ are in agreement with each other within our numerical accuracy of at most $10\,\%$ even though the computational methods are different.

\begin{figure}[t]
    \centering
    \includegraphics[width=0.45\textwidth]{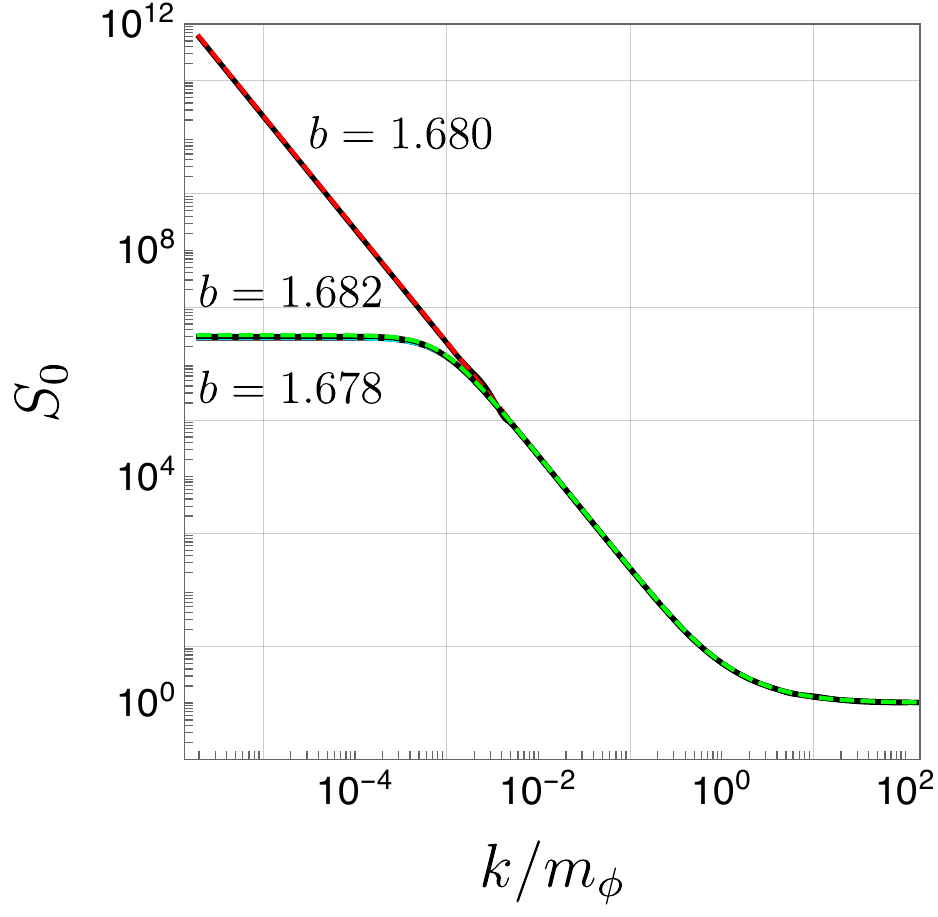}
    \includegraphics[width=0.45\textwidth]{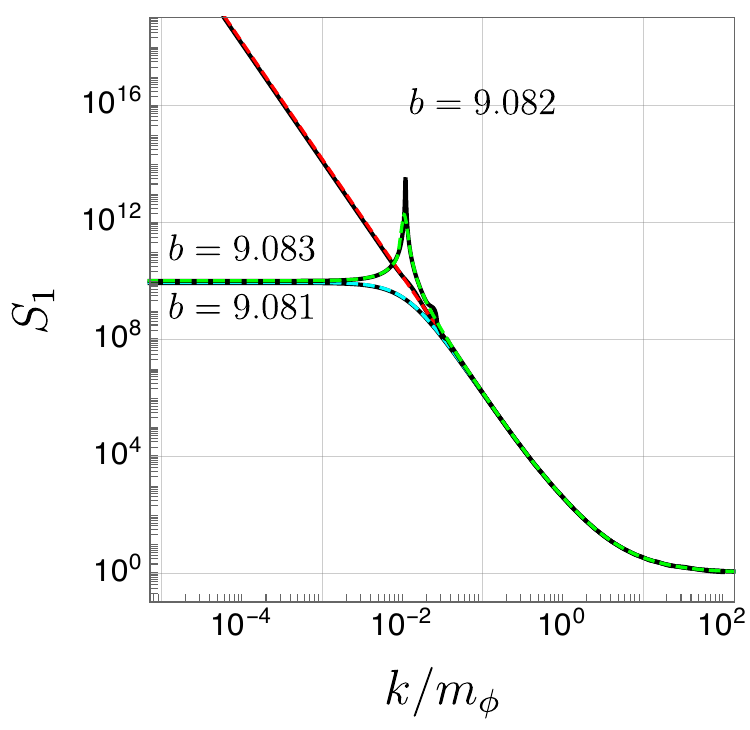}
    \caption{
        Same as \cref{fig:YukawaS}, but for the parameter $b$ near and on the first zero-energy resonance.
        Different line colors correspond to different choices of $b$ for $s$-wave ($p$-wave): $b = 1.680$ ($b = 9.082$) for on the first zero-energy resonance, $b = 1.678$/$b = 1.682$ ($b = 9.081$/$b = 9.083$) for below/above the first zero-energy resonance.
    }
    \label{fig:YukawaS2}
\end{figure}

Next, we discuss the Sommerfeld enhancement factors for $b$ near the first zero-energy resonances, which are marked as $\blacktriangle\,, \blacksquare$\,, and \tikzcircle{4pt} in \cref{fig:Yukawa_sl}.
The number of bound states is zero for the parameter $b$ below the zero-energy resonance and on the zero-energy resonance only for $s$-wave, and hence $F_\ell(k^2) = 1$.
Meanwhile, there is a bound state for the parameter $b$ above the zero-energy resonance and on the zero-energy resonance for the partial-wave with multipole $\ell \geq 1$. 
The rational function is given by $F_\ell(k^2) = k^2/(k^2+\kappa_b^2)$ for these parameters. 
As discussed in \cref{sec:ERT}, we can determine the pole $\kappa_b$ by the use of the effective range theory as far as $\kappa_b \lesssim |r_{e,\ell}|^{-1}$.
In other words, the pole determination by the use of the effective range theory is valid only when it is a shallow bound state.

We compare the Sommerfeld enhancement factors near the first zero-energy resonance, computed in two different ways, in \cref{fig:YukawaS2}.
These factors are in agreement with each other within our numerical accuracy of about $10\,\%$.
For $p$-wave Sommerfeld enhancement factor with $b = 9.081$, a spiky behavior appears at $k/m_\phi \simeq 10^{-2}$ due to a resonance as with the scattering cross section.
As mentioned in the previous section, this peak is correctly reproduced in our method even though we incorporate only the bound states in the rational function $F_\ell(k^2)$. 
In other words, the Omn\`es function is responsible for reproducing not only the branch cut of the amplitude but also the resonances that are closer to the physical axis in the second Riemann sheet.
It is worthwhile to understand how the resonances are incorporated in the Omn\`es function, but it is beyond the scope of this study. 

As in \cref{eq:OmnesFlowk}, we determine $F_\ell(k^2)$ so that $S_\ell$ is constant at small $k$ except for the zero-energy resonance. 
On the other hand, at the first zero-energy resonance, the factor diverges at small $k$ as $S_0 \sim k^{-2}$ and $S_\ell \sim k^{-4}$ for $\ell \geq 1$ in our formulation.
For the first zero-energy resonance of $s$-wave, there is only a virtual state, and hence $F_0(k^2) = 1$.
Meanwhile, for the first zero-energy resonance of higher partial-wave ($\ell \geq 1$), there exists a zero-energy bound state with $\kappa_b^2 = 0$, and hence $F_\ell(k^2) = 1$ again.
From the Levinson's theorem, we find the phase shift at $k = 0$ with our normalization $\delta_\ell(\infty) = 0$, $\delta_\ell(0) = \pi/2$ for the first zero-energy resonance of the $s$-wave and $\delta_\ell(0) = \pi$ for the first zero-energy resonance of the $p$-wave and higher partial-wave. 
$\omega_\ell (k^2)$ logarithmically diverges near $k^2 = 0$, and the Omn\`es function for the first zero-energy resonance at small $k$ results in 
\eqs{
    e^{\omega_0(k^2)} & \simeq \left( \frac{\Lambda^2}{k^2} \right)^{\delta_0(0)/\pi} = \frac{\Lambda}{k} \,, \qquad (s\text{-wave}) \,, \\
    e^{\omega_0(k^2)} & \simeq \left( \frac{\Lambda^2}{k^2} \right)^{\delta_0(0)/\pi} = \frac{\Lambda^2}{k^2} \,, \qquad (\text{partial-wave with } \ell \geq 1) \,, 
    \label{eq:smallk2}
}
where $\Lambda \sim r_{e,\ell}^{-1}$.
Combining them, we get the small-$k$ behavior of the Sommerfeld enhancement factor at the zero-energy resonance as shown above.
\cref{fig:YukawaS2} shows that our formula reproduces the small-$k$ behavior of the factor via the direct computation.
Since we choose the function $F_\ell(k^2)$ so that the Sommerfeld enhancement factor is constant at small $k$ except for the zero-energy resonances, we obtain the constant value for the factor for the parameter near the first zero-energy resonance.
Below the first zero-energy resonance, $F_\ell(k^2) = 1$ and the Omn\`es function is constant due to $\delta_\ell(0) = 0$.
For $b$ above the first zero-energy resonance, $F_\ell(k^2) = k^2/(k^2+\kappa_b^2)$ and the Omn\`es function scales as $e^{\omega_\ell(k^2)} \simeq \Lambda^2/k^2$.
As with the case on the first zero-energy resonance, our formula reproduces the small-$k$ behavior of the factor via the direct computation.
In particular, our choice of $\kappa_b$ by the use of effective range theory well reproduces the direct computation for the parameter above the first zero-energy resonance.%
\footnote{
    In the text, we assume that $\delta_\ell(k^2)$ is given. 
    If $\delta_\ell(k^2)$ is not available but only $a_\ell$ and $r_{e,\ell}$ are known, we may still determine the small-$k$ scaling of $S_\ell$ near the zero-energy resonance, $|a_\ell| \gg |r_{e,\ell}|$, in the same way. 
    By restricting $\delta_\ell(k^2) \in [0 , \pi)$, $\delta_\ell(k^2)$ is uniquely determined at small $k$.
    In this way, we miss the deep bound states if exist (remember the Levinson's theorem), but they only affect the normalization but not scaling of $S_\ell$ below these bound states.
}

Before closing this section, we comment on the partial-wave unitarity of the Sommerfeld-enhanced cross section on the zero-energy resonance.
Since the hard annihilation cross section with multipole $\ell$ scales as $(\sigma_{\mathrm{ann},\ell} v)_0 \propto v^{2 \ell}$ at small velocity, the velocity-weighted cross section $\sigma_{\mathrm{ann},\ell} v$ scales as $v^{-2}$ for $s$-wave and $v^{2 \ell - 4}$ for higher partial-waves $\ell \geq 1$. 
Therefore, the annihilation cross section violates the partial-wave unitarity $\sigma_\mathrm{ann} v \leq 4 \pi (2 \ell + 1)/m_\chi^2 v$ at small velocity for $s$- and $p$-waves.
The annihilation cross section should be regularized for these partial waves to restore the partial-wave unitarity by taking into account the short-range interaction~\cite{Blum:2016nrz}.

\section{Conclusion \label{sec:conclusion}}

In this study, we have addressed the correlation between the resonant DM self-scattering and the large Sommerfeld enhancement factor for the DM annihilation.
We have directly formulated the relation between the scattering phase shift and the Sommerfeld enhancement factor by the use of Watson's theorem.
This is one of powerful theorems of quantum-mechanical scattering process: the discontinuity of the annihilation amplitude is determined by the scattering phase shift. 
The form of the amplitude is given by Omn\`es solution, which reproduces the pole structure and the branch cut of the amplitude.
We have computed the Sommerfeld enhancement factor by the use of Omn\`es solution for a given phase shift for attractive Yukawa potential mediated by a light mediator. 
Our formulation has accurately reproduced the Sommerfeld enhancement factor, which is computed directly by solving the Schr\"odinger equation, with suitable choices of a function $F_\ell(k^2)$ responsible for the pole structure of the amplitude.
Moreover, we can use the effective range theory to determine the poles $\kappa_b$ when the self-scattering is close to the zero-energy resonance.
If the velocity dependence of the DM self-scattering cross section is confirmed by observations, we can also determine the Sommerfeld enhancement factor by the use of our formalism, which is important for DM indirect searches.

\section*{Acknowledgements}
A. K. acknowledges partial support from Grant-in-Aid for Scientific Research from the Ministry of Education, Culture, Sports, Science, and Technology (MEXT), Japan, 18K13535 and 19H04609; from World Premier International Research Center Initiative (WPI), MEXT, Japan; from Norwegian Financial Mechanism for years 2014-2021, grant nr 2019/34/H/ST2/00707; and from National Science Centre, Poland, grant 2017/26/E/ST2/00135 and DEC-2018/31/B/ST2/02283.
The work of T. K. is supported in part by the National Science Foundation of China under Grant Nos. 11675002, 11635001, 11725520, 12235001, and 12250410248.

\appendix

\section{Hulth\'en Potential \label{app:Hulthen}}

In this appendix, we consider a two-body system under the attractive Hulth\'en potential
\eqs{
    V(r) = - \frac{\alpha m_\ast e^{- m_\ast r}}{1-e^{-m_\ast r}} \,.
    \label{eq:HulthenP}
}
This potential approximates the Yukawa potential: this scales as $1/r$ at short distances, and the force mediated by the potential is screened at large distances. 
$m_\ast$ denotes the screening mass, and hence we assume that $m_\ast$ is related to $m_\phi$ as $m_\ast = \kappa m_\phi$ with an $\mathcal{O}(1)$ coefficient $\kappa$.%
\footnote{
    The coefficient $\kappa$ is chosen to be $\kappa = \sqrt{2 \zeta(3)} = 1.55 \dots $ by equating the Born scattering cross section with Hulth\'en and Yukawa potentials~\cite{Tulin:2013teo}.
    As for the Sommerfeld enhancement factor, a similar prescription provides $\kappa = \zeta(2) = \pi^2/6 = 1.64 \dots $. 
    However, the positions of the resonant enhancement for these processes should be agree with each other as seen in the case of Yukawa potential, and hence the coefficient must be identical.
}
The Schr\"odinger equation with the potential is analytically solvable for $s$-wave. 
For the higher partial-waves ($\ell \geq 1$), however modification of the centrifugal term allows us to find the analytical solution~\cite{Greene:1976a}:
\eqs{
    \frac{1}{r^2} \frac{d}{d r} \left( r^2 \frac{d R_{k,\ell}}{d r} \right) + \left( k^2 - \ell(\ell+1) \frac{m_\ast^2 e^{-m_\ast r}}{(1-e^{-m_\ast r})^2}  - 2 \mu V(r) \right) R_{k,\ell} = 0 \,.
    \label{eq:rad_Hul}
}
The modified centrifugal term reproduces $\ell(\ell+1)/r^2$ at short distance $m_\ast r \ll 1$.

\begin{figure}[t]
    \centering
    \includegraphics[width=0.45\textwidth]{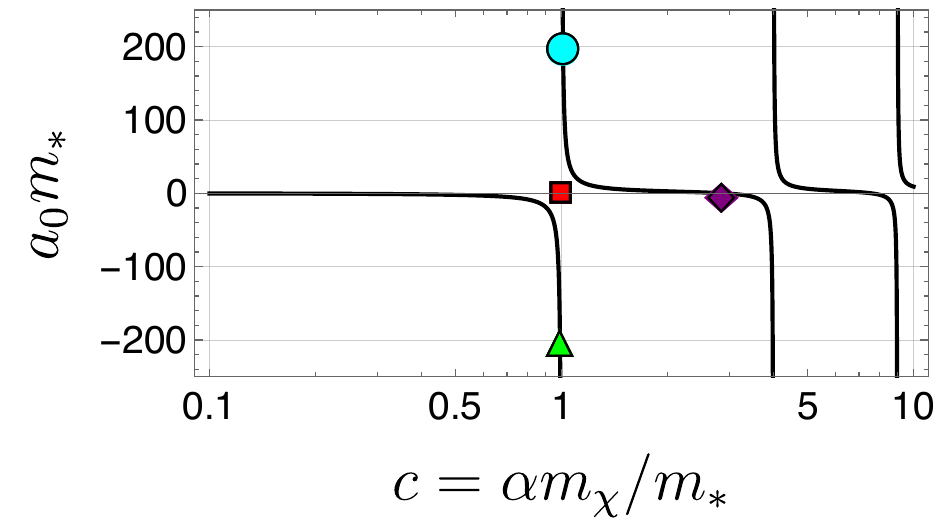}
    \includegraphics[width=0.45\textwidth]{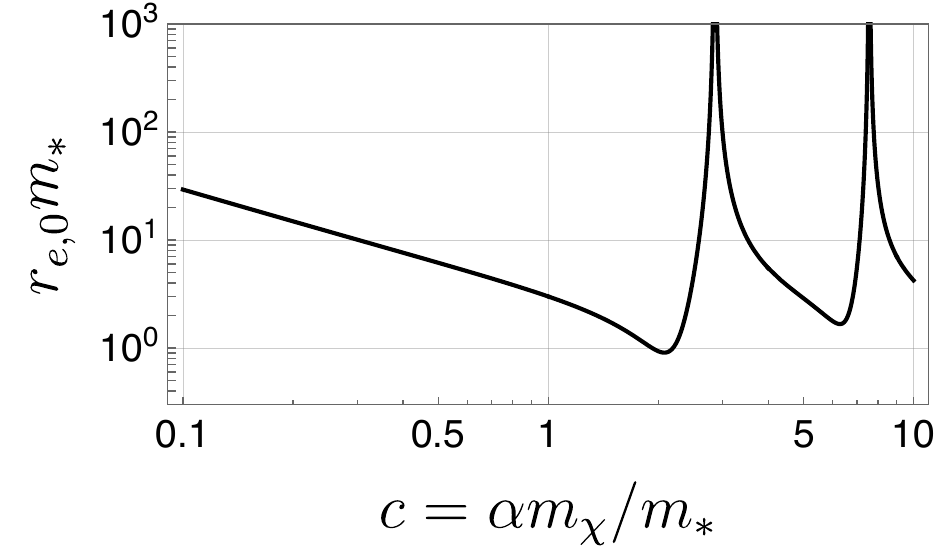}
    \caption{
        Scattering length (\textit{left}) and effective range (\textit{right}) for $s$-wave scattering in the Hulth\'en potential as a function of $c$ in unit of $m_\ast$.
        Since the scattering length diverges at $b = 1.0$, we put the mark for the point on the $c$-axis. 
    }
    \label{fig:Hulthen_sl}
\end{figure}

We can find the analytic form of the phase shift under the Hulth\'en potential.
Defining $a \equiv v/2 \alpha$ and $c \equiv \alpha m_\chi/m_\ast$ and introducing a new variable $t \equiv 1 - e^{-m_\ast r}$, the analytic solution of the radial wave-function is given by~\cite{Cassel:2009wt} 
\eqs{
    R_{k,\ell}(r)
    = \frac{t^{\ell+1}}{2r} \frac{e^{-ikr}}{\Gamma(2\ell+2)}
    \left| \frac{\Gamma(\lambda^\ell_+)\Gamma(\lambda^\ell_-)}{\Gamma(\lambda^\ell_+ +\lambda^\ell_- - 2 - 2 \ell )} \right| 
    {}_2F_1 (\lambda^-,\lambda^+;2\ell+2;t) \,,
}
where ${}_2F_1$ denotes the hypergeometric function, and $\lambda^\ell_\pm \equiv 1 + \ell + i a c \pm \sqrt{c - a^2 c^2}$, and $\Gamma(x)$ is the gamma function.
In the large distance limit ($r \to \infty$, \textit{i.e.}, $t \to 1$), ${}_2F_1 (\lambda^-,\lambda^+;2\ell+2;t)$ behaves as 
\eqs{
    & {}_2F_1 (\lambda^-,\lambda^+;2\ell+2;t) \\
    & \to \frac{\Gamma(2\ell+2) \Gamma(2 + 2 \ell - \lambda^\ell_+ - \lambda^\ell_-)}{\Gamma(2 + 2 \ell - \lambda^\ell_+)\Gamma(2 + 2 \ell - \lambda^\ell_-)}
    + e^{2 i k r} \frac{\Gamma(2\ell+2) \Gamma(\lambda^\ell_+ +\lambda^\ell_- - 2 - 2 \ell )}{\Gamma(\lambda^\ell_+)\Gamma(\lambda^\ell_-)} \,.
}
Hence, the radial wave-function $\chi_{k,\ell}(r) \equiv r R_{k,\ell}(r)$ has the asymptotic behavior as
\eqs{
    \chi_{k,\ell}(r) & \to \frac12 
    \left| \frac{\Gamma(\lambda^\ell_+)\Gamma(\lambda^\ell_-)}{\Gamma(\lambda^\ell_+ +\lambda^\ell_- - 2 - 2 \ell )} \right| \\
    & \times \left[ \frac{\Gamma(\lambda^\ell_+ +\lambda^\ell_- - 2 - 2 \ell )}{\Gamma(\lambda^\ell_+)\Gamma(\lambda^\ell_-)} e^{ikr}
    + \frac{\Gamma(2 + 2 \ell - \lambda^\ell_ + -\lambda^\ell_-)}{\Gamma(2 + 2 \ell - \lambda^\ell_+)\Gamma(2 + 2 \ell - \lambda^\ell_-)} e^{-ikr} 
    \right] \,,
    \label{eq:asymptotic_chiH}
}
Matching the asymptotic behavior of $\chi_{k,\ell}(r)$ to the asymptotic form of the radial wave-function, the phase shift is analytically obtained. 
The phase shift is computed as the difference of the phases with and without the potential term.
\eqs{
    \delta_\ell = \arg \left( \frac{i \Gamma(\lambda^\ell_+ + \lambda^\ell_- - 2 - 2 \ell)}{\Gamma(\lambda^\ell_+)\Gamma(\lambda^\ell_-)} \right) 
    - \left.  
        \arg \left( \frac{i \Gamma(\lambda^\ell_+ + \lambda^\ell_- - 2 - 2 \ell)}{\Gamma(\lambda^\ell_+)\Gamma(\lambda^\ell_-)} \right) 
    \right|_{\substack{c \to 0 \\ac = k/m_\ast: \mathrm{fixed}}}\,.
    \label{eq:HulthenPh}
}
Here, the second term is in agreement with $ + \ell \pi/2$ in the limit of $k \to \infty$, which is an inherent phase for the scattering with partial-wave $\ell$ in the absence of the potential term when the centrifugal term is not modified.
This $s$-wave phase shift gives the scattering length $a_0$ and the effective range $r_{e,0}$ as 
\eqs{
    a_0 & = \frac{1}{m_\ast} \left[ \psi^{(0)}(1 + \sqrt{c}) + \psi^{(0)}(1 - \sqrt{c}) + 2 \gamma_E \right] \,, \\
    r_{e,0} & = \frac{2 a_0}{3} 
    - \frac{1}{m_\ast^3 a_0^2 \sqrt{c}} \left[ \psi^{(1)}(1 + \sqrt{c}) - \psi^{(1)}(1 - \sqrt{c})\right] \\
    & \quad - \frac{1}{3 m_\ast^3 a_0^2} \left[ \psi^{(2)}(1 + \sqrt{c}) + \psi^{(2)}(1 - \sqrt{c}) + 16 \zeta(3) \right] \,.
}
Here, $\psi^{(n)}(x)$ is the polygamma function of order $n$, and $\gamma_E = 0.577\dots$ denotes the Euler--Mascheroni constant.
The scattering length diverges at $c = n^2$ with integers $n$. 
We illustrate the behavior of $a_0$ and $r_{e,0}$ as a function of $c$ in \cref{fig:Hulthen_sl}.

For the $p$- and higher partial-waves scattering, we do not obtain the scattering length and the effective range that correctly reproduce those for Yukawa potential due to the modification of the centrifugal term of the Schr\"odinger equation. 
Indeed, at small $k$, the phase shift $\delta_\ell$ is expanded as 
\eqs{
    \delta_\ell
    \simeq - k \left[ \psi^{(0)}(1 + \ell + \sqrt{c}) + \psi^{(0)}(1 + \ell - \sqrt{c}) - 2 \psi^{(0)}(1 + \ell )\right]\,, 
}
and hence $k^{2 \ell+1} \cot \delta_\ell$ starts from $k^{2 \ell}$ terms at small $k$.
Specifically, we obtain the effective-range theory parameters for the higher partial-waves as,
\eqs{
    a_1^{-1} & = 0 \,, \qquad 
    r_{e, 1} = - \frac{1}{2 m_\ast} \left[ \psi^{(0)}(2 + \sqrt{c}) + \psi^{(0)}(2 - \sqrt{c}) + 2 \gamma_E \right] \,, 
}
for $p$-wave, and $a_\ell^{-1} = 0$ and $r_{e, \ell}^{-1} = 0$ for the partial-waves with $\ell \geq 2$.
Since the modified centrifugal potential rapidly decreases as $e^{-m_\ast r}$ not $r^{-2}$ at large $r$, the phase shift $\delta_\ell$ behaves like $s$-wave phase shift rather than the higher partial-wave.

We also understand from \cref{eq:asymptotic_chiH} that the $s$-wave zero-energy state is a virtual level.
For the bound state solution above the first zero-energy resonance, $k = i \kappa_b = i (c-1)m_\ast/2$, and the radial wave-function is given by
\eqs{
    \chi_{k,0}(r) & \to \frac12
    \left| \frac{\Gamma(\tilde \lambda^0_+)\Gamma(\tilde \lambda^0_-)}{\Gamma(\tilde \lambda^0_+ +\tilde \lambda^0_- - 2 )} \right| \frac{ \Gamma(\tilde \lambda^0_+ +\tilde \lambda^0_- - 2 )}{\Gamma(\tilde \lambda^0_+)\Gamma(\tilde \lambda^0_-)} e^{-\kappa_b r} 
    = \frac12 e^{-\kappa_b r} \,.
}
Here, $\tilde \lambda^0_\pm \equiv 1 - \kappa_b \pm \sqrt{c + \kappa_b^2}$ is a real constant.
It indicates that the zero-energy state ($\kappa_b = 0$) has a constant wave-function, and hence it is a virtual level. 
Even for the higher partial-waves ($\ell \geq 1$), the zero-energy states ($\kappa_b = 0$) have constant wave-functions and are virtual levels unlike the case of the Yukawa potential. 

We can find the analytic form for the enhancement factor of annihilation under the Hulth\'en potential with the same quantities $\lambda^\ell_\pm$~\cite{Cassel:2009wt,Slatyer:2009vg,Feng:2010zp}. 
In the short distance limit ($r \to 0$, \textit{i.e.}, $t \to 0$), the function ${}_2F_1 (\lambda^-,\lambda^+;2\ell+2;t) \to 1 $, and hence the radial wave-function behaves as
\eqs{
    \chi_{k,\ell}(r) & \to 
    \frac{(m_\ast r)^{\ell+1}}{2 \Gamma(2 \ell + 2)} \left| \frac{\Gamma(\lambda^\ell_+)\Gamma(\lambda^\ell_-)}{\Gamma(\lambda^\ell_+ +\lambda^\ell_- - 2 - 2 \ell )} \right| \,.
    \label{eq:asymptotic_chiH0}
}
Using the asymptotic behavior of $\chi_{k,\ell}$, given by \cref{eq:chi_at_0}, we find a constant $B_\ell$ which describes the distortion of the wave-function near the origin.
\eqs{
    B_\ell = \frac{1}{(ac)^{\ell+1}}\frac{(2 \ell+1)!!}{2 \Gamma(2 \ell + 2)} \left| \frac{\Gamma(\lambda^\ell_+)\Gamma(\lambda^\ell_-)}{\Gamma(\lambda^\ell_+ +\lambda^\ell_- - 2 - 2 \ell )} \right| \,.
}
Due to the modification of the centrifugal term, the wave-function without the potential term is also distorted at the origin. 
Hence, the Sommerfeld enhancement factor is computed as the ratio of $|B_\ell|^2$ with and without the potential term:
\eqs{
    S_\ell = \frac{|B_\ell|^2}{|B_\ell(c \to 0; ac = k/m_\ast)|^2} 
    = \left| \frac{\Gamma(\lambda^\ell_+)\Gamma(\lambda^\ell_-)}{\Gamma(\lambda^\ell_ + +\lambda^\ell_- - 1 - \ell)\ell!} \right|^2 \,.
    \label{eq:HulthenS}
}

\begin{figure}[t]
    \centering
    \includegraphics[width=0.45\textwidth]{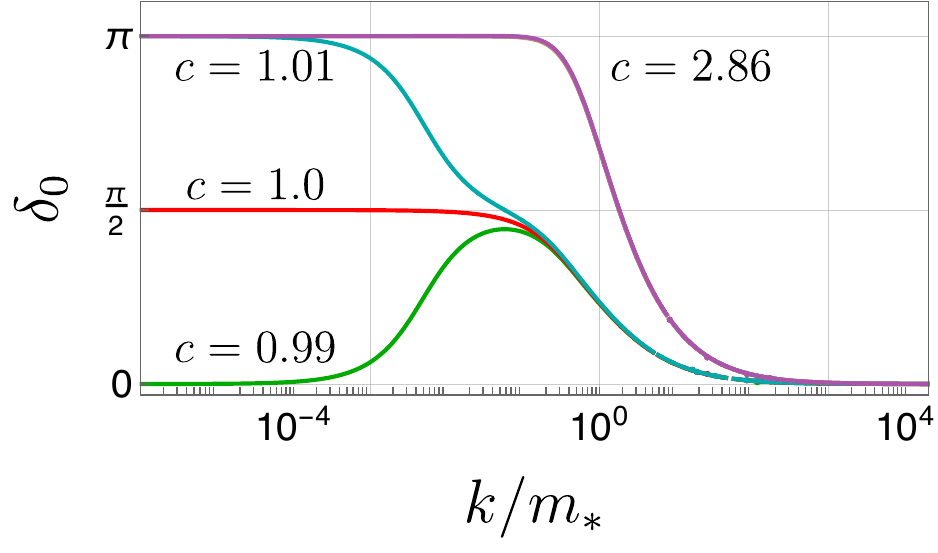}
    \includegraphics[width=0.45\textwidth]{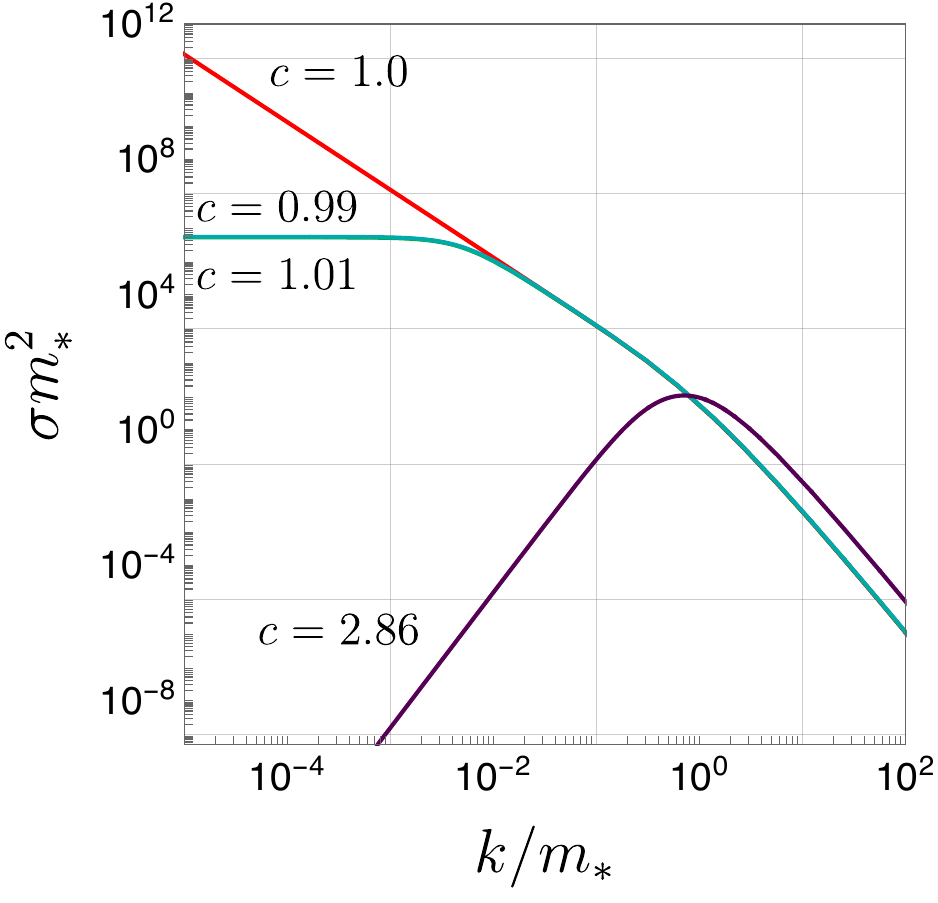}
    \caption{
        Phase shift and cross section under the Hulth\'en potential. 
        (\textit{left}): 
        Phase shift of $s$-wave scattering for given $c$ as a function of $k/m_\ast$. 
        (\textit{right}): 
        Cross section in unit of $m_\ast$ for given $c$ as a function of $k/m_\ast$. 
    }
    \label{fig:Hulthen}
\end{figure}

\begin{figure}[t]
    \centering
    \includegraphics[width=0.45\textwidth]{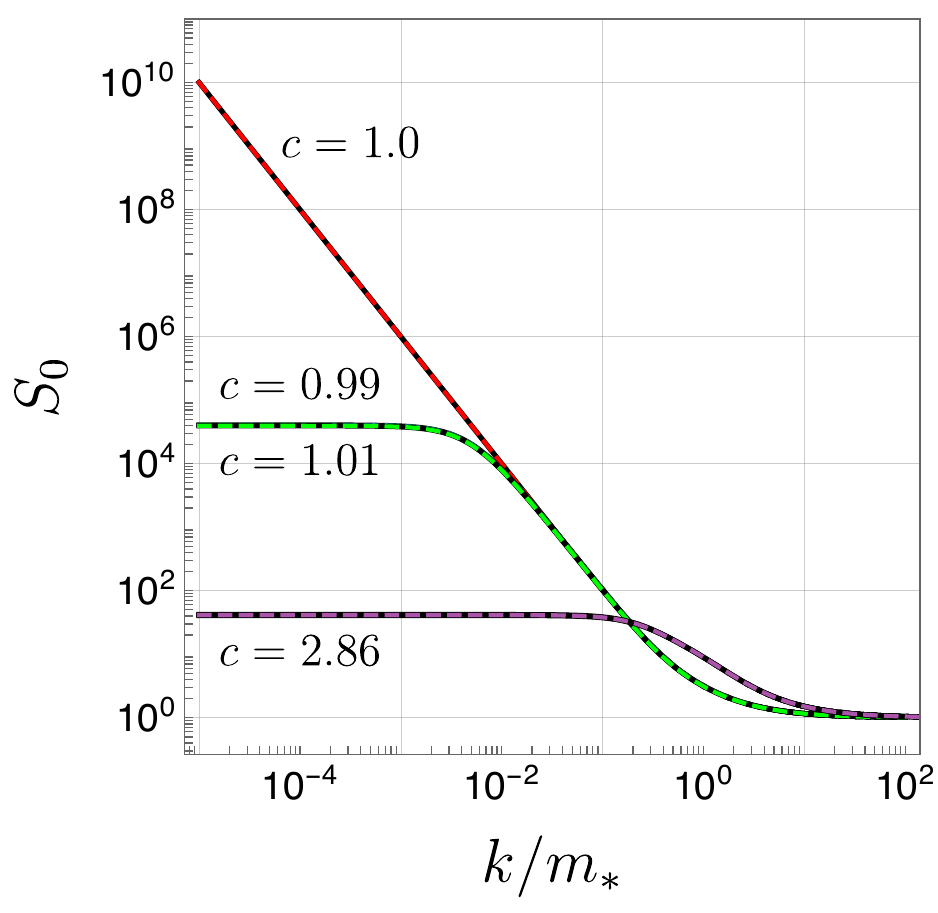}
    \includegraphics[width=0.45\textwidth]{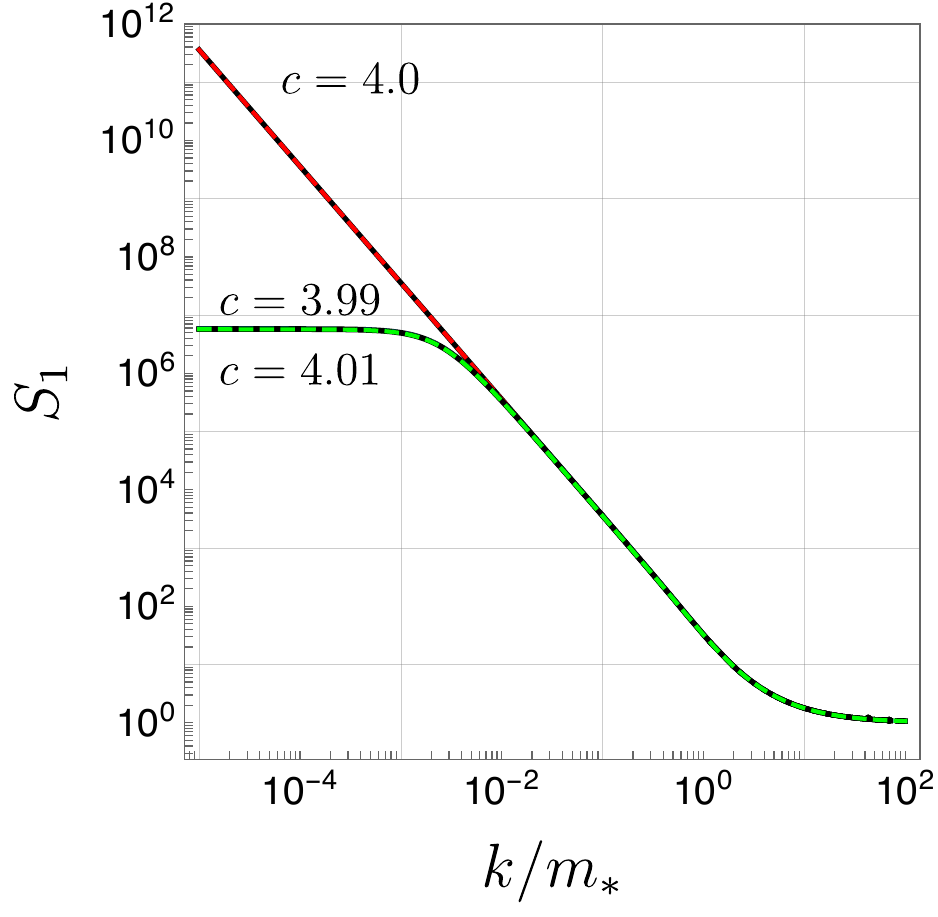}
    \caption{
        Sommerfeld enhancement factors for given $c$ as a function of $k/m_\ast$.
        $S_\ell$ from the analytic form is shown as black-solid lines, while $S_\ell$ from the phase shift and Watson's theorem is shown as colored-dashed lines.
    }
    \label{fig:HulthenS}
\end{figure}

We have analytic forms of the phase shift and the Sommerfeld enhancement factor for the scattering and annihilation processes under the Hulth\'en potential, as shown in \cref{eq:HulthenPh,eq:HulthenS}.
As discussed in the text, for given phase shift, we can compute the amplitude of the DM annihilation through the Omn\`es solution, given by \cref{eq:OmnesF}.
We now compare the $s$-wave and $p$-wave Sommerfeld enhancement factors computed in the two different ways.
For $s$-wave, we take four reference points, which are marked in \cref{fig:Hulthen_sl}: $c = 0.99$ ($\blacktriangle$) just below the first zero-energy resonance, $c = 1.0$ ($\blacksquare$) on the first zero-energy resonance, $c = 1.01$ (\tikzcircle{4pt}) just above the first zero-energy resonance, and $c = 2.86$ ($\blacklozenge$) on the first anti-resonance.
For $p$-wave, we take three reference points: $c = 3.99$ just below the first zero-energy resonance, $c = 4.0$ on the first zero-energy resonance, and $c = 4.01$ just above the first zero-energy resonance.
We summarize the phase shift and the cross section for $s$-wave in \cref{fig:Hulthen}, and comparison of the Sommerfeld enhancement factors computed in the two different ways in \cref{fig:HulthenS}.
These factors are numerically in agreement with each other with an accuracy of a few per mille.
Compared to the Yukawa potential, we have the analytic formulae for the phase shift and the Sommerfeld enhancement factor, and hence these factors are identical in a high precision.

The behavior of the phase shift and the cross section is understood in a similar way to the Yukawa potential.
For all of the reference points, we find good agreement of $S_\ell$ given by the Omn\`es solution \cref{eq:OmnesF} (colored-dashed lines) with the analytic form (black-solid lines).
We would emphasize that, for the Hulth\'en potential, we can find the pole position even for the parameters far from the first zero-energy resonance, in particular the first anti-resonance, thanks to the analytic form of the wave-function.
The bound-state solution should be $\chi_{k,\ell} \propto e ^{-\kappa_b r}$ when we take $k = i \kappa_b$, and thus a coefficient of the second term of \cref{eq:asymptotic_chiH} vanishes.  
We find the pole to be $\kappa_b = (c-1)m_\ast/2$ for $s$-wave and $\kappa_b = (c-4)m_\ast/4$ for $p$-wave above the first zero-energy resonance.
Even though the effective-range theory parameters are not agree with that of the Yukawa potential for $p$- and higher-waves, the Sommerfeld enhancement factors for $p$-wave are accurately in agreement with each other (the right panel of \cref{fig:HulthenS}).

In this appendix, we numerically investigate the relation between the phase shift and the Sommerfeld enhancement factor under the Hulth\'en potential. 
This indicate that one may have an analytic relation of the gamma function providing a relation between \cref{eq:HulthenPh} and \cref{eq:HulthenS}. 
However, we could not find such relation in the literature. 

\section{Levinson's Theorem \label{app:Levinson}}

As discussed in \cref{sec:Watson}, there is a theorem that relates the behavior of the phase shift and the number of bound states. 
The original article \cite{Levinson:407337} provides a rigorous proof of the theorem.
We give a na\"ive sketch of a proof for a simple system, which is given in Refs.~\cite{Wellner:1964,Weinberg2015lectures}. 
We assume that the system has a central-force potential and that the system is enclosed in a large sphere of radius $R$. 
As far as the potential rapidly vanishes at large radius $r$, the scattering state for given angular momentum $\ell$ is approximately proportional to $\sin(k r - \ell \pi/2 + \delta_\ell(k))$. 
The boundary condition at $r = R$ requires that the wave-function vanishes and the scattering state has a discretized momentum $k_n$ ($n$ denoting an arbitrary integer that gives positive $k_n R$). 
\eqs{
    k_n R - \frac{\ell \pi}{2} + \delta_\ell(k_n) = n \pi \,.
}
Let $n'$ be the minimum integer that gives a positive $k_{n'} R$.
For a given momentum $k_{n}$, the difference of the phase shift is written as 
\eqs{
    \frac{1}{\pi} \left[ \delta_\ell(k_n) - \delta_\ell(k_{n'}) \right] + \frac{1}{\pi} (k_n - k_{n'}) R = N_\ell(k_n) \,.
}
Here, $N_\ell (k_n)$ denotes the number of the scattering states with the momentum in the range of $k_{n'} \leq k < k_n$. 
In the absence of the potential, the phase shift vanishes and $(k_n - k_{n'}) R/\pi$ gives the number of the scattering states with the momentum in a range of $k_{n'} \leq k \leq k_n$.
Therefore, the number of the scattering states can change due to the presence of the potential, and its variation is given by 
\eqs{
    \Delta N_\ell(k_n) = \frac{1}{\pi} \left[ \delta_\ell(k_n) - \delta_\ell(0) \right] \,.
}
Here, we take $k_{n'} \to 0$ for a sufficiently large radius $R$. 
As $k_n \to \infty$, this gives the total change of the number of the scattering states. 
Meanwhile, the total number of states is unchanged as far as we gradually turn on the potential.
Some of the scattering states with the energy $E > 0$ for $V = 0$ becomes the bound state with the negative energy in the presence of $V$.
Therefore, a portion of the scattering states is converted into the bound state, and the number of the bound state is given by
\eqs{
    N_b = - \Delta N_\ell(\infty) 
    = \frac{1}{\pi} \left[ \delta_\ell(0) - \delta_\ell(\infty) \right]\,.
}
This is known as the Levinson's theorem that gives a relation between the behavior of the phase shift and the number of bound states.
We note that there is a subtlety related to zero-energy bound states and zero-energy virtual levels. 
As shown in \cref{eq:Levinson}, there is an extra term only when we consider the $s$-wave zero-energy resonance.

\section{d-wave case \label{app:dwave}}

\begin{figure}[t]
    \centering
    \includegraphics[width=0.45\textwidth]{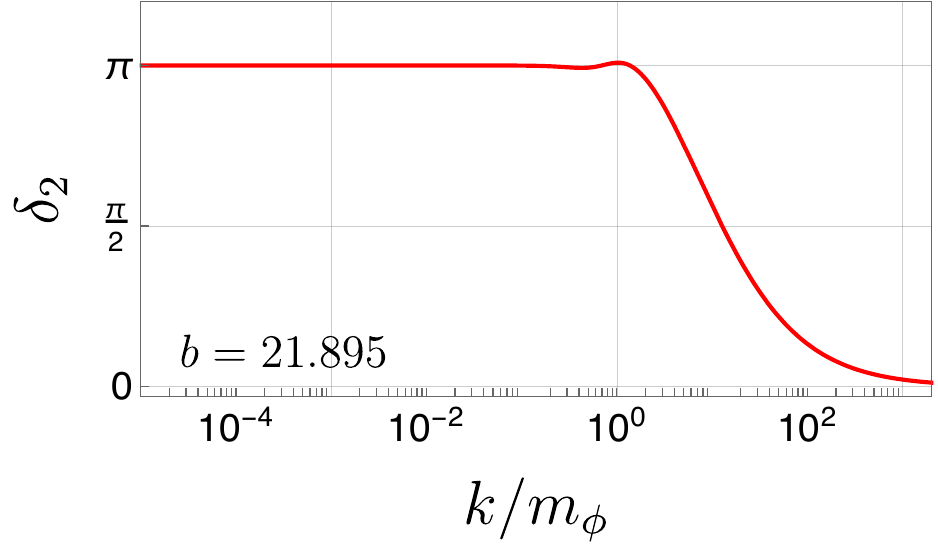}
    \includegraphics[width=0.45\textwidth]{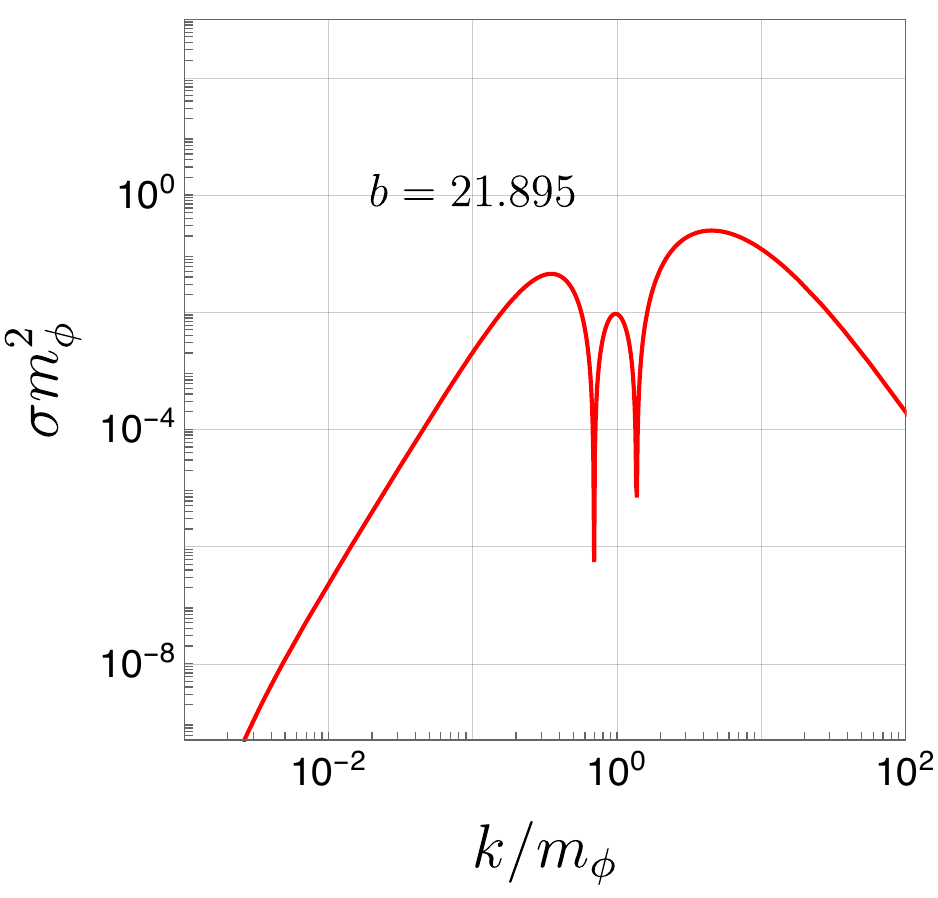}
    \caption{
        Same as \cref{fig:Yukawa}, but for $d$-wave phase shift and cross section on the first zero-energy resonance. 
    }
    \label{fig:Yukawad}
\end{figure}

In the text, we discuss the agreement of the Sommerfeld enhancement factors for $s$-wave and $p$-wave in two different computational methods. 
We also demonstrate that the use of the Omn\`es solution numerically works even for the $d$-wave case.

We show the phase shift $\delta_2(k)$ and the cross section in the unit of $m_\phi$ on the first zero-energy resonance ($b = 21.895$) in \cref{fig:Yukawad}.
The phase shift scales as $\cot \delta_2 \simeq (- r_{e,2} k)^{-3}$ at small $k$, and hence the cross section scales as $k^4$ at small $k$.
We find that the phase shift crosses $\pi$ around $k \sim m_\phi$, and hence the scattering cross section has dips around there. 
We do not find the physical reason for that.

\begin{figure}[t]
    \centering
    \includegraphics[width=0.45\textwidth]{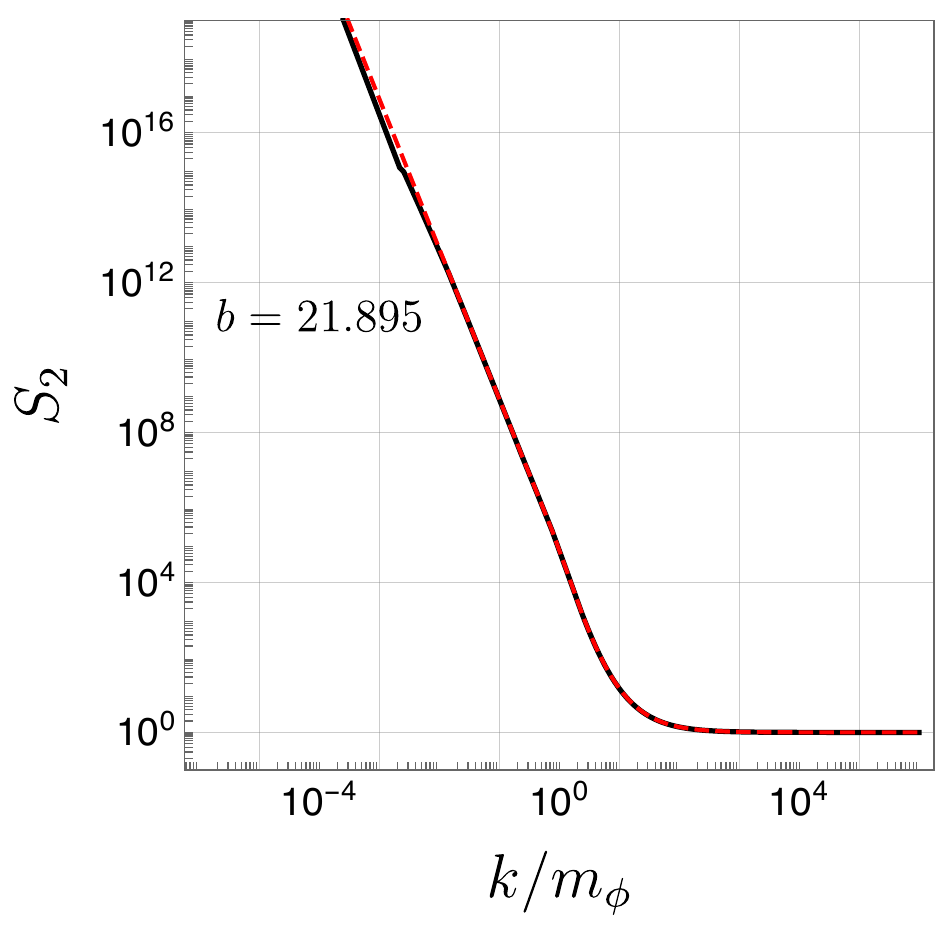}
    \caption{
        Same as \cref{fig:YukawaS}, but for $d$-wave Sommerfeld enhancement factors on the first zero-energy resonance as a function of $k/m_\phi$.
    }
    \label{fig:YukawaellS}
\end{figure}
\cref{fig:YukawaellS} shows the Sommerfeld enhancement factors computed in two different ways as discussed in the text.
Since the phase shift approaches $\delta_2 \to \pi$ at small $k$, the Omn\`es function is logarithmically diverges as $e^{\omega_2(k^2)} \simeq \Lambda^2/k^2$ with $\Lambda \sim r_{e,2}^{-1}$.
Even for the $d$-wave, we find that these factors in two ways are in agreement with each other.

\bibliographystyle{utphys}
\bibliography{ref}

\providecommand{\href}[2]{#2}\begingroup\raggedright\begin{thebibliography}{10}

\bibitem{Spergel:1999mh}
D.~N. Spergel and P.~J. Steinhardt, ``{Observational evidence for
  selfinteracting cold dark matter},''
  \href{http://dx.doi.org/10.1103/PhysRevLett.84.3760}{{\em Phys. Rev. Lett.}
  {\bfseries 84} (2000) 3760--3763},
  \href{http://arxiv.org/abs/astro-ph/9909386}{{\ttfamily
  arXiv:astro-ph/9909386}}.

\bibitem{Tulin:2017ara}
S.~Tulin and H.-B. Yu, ``{Dark Matter Self-interactions and Small Scale
  Structure},'' \href{http://dx.doi.org/10.1016/j.physrep.2017.11.004}{{\em
  Phys. Rept.} {\bfseries 730} (2018) 1--57},
  \href{http://arxiv.org/abs/1705.02358}{{\ttfamily arXiv:1705.02358
  [hep-ph]}}.

\bibitem{Kaplinghat:2015aga}
M.~Kaplinghat, S.~Tulin, and H.-B. Yu, ``{Dark Matter Halos as Particle
  Colliders: Unified Solution to Small-Scale Structure Puzzles from Dwarfs to
  Clusters},'' \href{http://dx.doi.org/10.1103/PhysRevLett.116.041302}{{\em
  Phys. Rev. Lett.} {\bfseries 116} no.~4, (2016) 041302},
  \href{http://arxiv.org/abs/1508.03339}{{\ttfamily arXiv:1508.03339
  [astro-ph.CO]}}.

\bibitem{Newman:2012nw}
A.~B. Newman, T.~Treu, R.~S. Ellis, and D.~J. Sand, ``{The Density Profiles of
  Massive, Relaxed Galaxy Clusters: II. Separating Luminous and Dark Matter in
  Cluster Cores},'' \href{http://dx.doi.org/10.1088/0004-637X/765/1/25}{{\em
  Astrophys. J.} {\bfseries 765} (2013) 25},
  \href{http://arxiv.org/abs/1209.1392}{{\ttfamily arXiv:1209.1392
  [astro-ph.CO]}}.

\bibitem{Randall:2008ppe}
S.~W. Randall, M.~Markevitch, D.~Clowe, A.~H. Gonzalez, and M.~Bradac,
  ``{Constraints on the Self-Interaction Cross-Section of Dark Matter from
  Numerical Simulations of the Merging Galaxy Cluster 1E 0657-56},''
  \href{http://dx.doi.org/10.1086/587859}{{\em Astrophys. J.} {\bfseries 679}
  (2008) 1173--1180}, \href{http://arxiv.org/abs/0704.0261}{{\ttfamily
  arXiv:0704.0261 [astro-ph]}}.

\bibitem{Harvey:2018uwf}
D.~Harvey, A.~Robertson, R.~Massey, and I.~G. McCarthy, ``{Observable tests of
  self-interacting dark matter in galaxy clusters: BCG wobbles in a constant
  density core},'' \href{http://dx.doi.org/10.1093/mnras/stz1816}{{\em Mon.
  Not. Roy. Astron. Soc.} {\bfseries 488} no.~2, (2019) 1572--1579},
  \href{http://arxiv.org/abs/1812.06981}{{\ttfamily arXiv:1812.06981
  [astro-ph.CO]}}.

\bibitem{Kamada:2016euw}
A.~Kamada, M.~Kaplinghat, A.~B. Pace, and H.-B. Yu, ``{How the Self-Interacting
  Dark Matter Model Explains the Diverse Galactic Rotation Curves},''
  \href{http://dx.doi.org/10.1103/PhysRevLett.119.111102}{{\em Phys. Rev.
  Lett.} {\bfseries 119} no.~11, (2017) 111102},
  \href{http://arxiv.org/abs/1611.02716}{{\ttfamily arXiv:1611.02716
  [astro-ph.GA]}}.

\bibitem{Ren:2018jpt}
T.~Ren, A.~Kwa, M.~Kaplinghat, and H.-B. Yu, ``{Reconciling the Diversity and
  Uniformity of Galactic Rotation Curves with Self-Interacting Dark Matter},''
  \href{http://dx.doi.org/10.1103/PhysRevX.9.031020}{{\em Phys. Rev. X}
  {\bfseries 9} no.~3, (2019) 031020},
  \href{http://arxiv.org/abs/1808.05695}{{\ttfamily arXiv:1808.05695
  [astro-ph.GA]}}.

\bibitem{Valli:2017ktb}
M.~Valli and H.-B. Yu, ``{Dark matter self-interactions from the internal
  dynamics of dwarf spheroidals},''
  \href{http://dx.doi.org/10.1038/s41550-018-0560-7}{{\em Nature Astron.}
  {\bfseries 2} (2018) 907--912},
  \href{http://arxiv.org/abs/1711.03502}{{\ttfamily arXiv:1711.03502
  [astro-ph.GA]}}.

\bibitem{Correa:2020qam}
C.~A. Correa, ``{Constraining velocity-dependent self-interacting dark matter
  with the Milky Way\textquoteright{}s dwarf spheroidal galaxies},''
  \href{http://dx.doi.org/10.1093/mnras/stab506}{{\em Mon. Not. Roy. Astron.
  Soc.} {\bfseries 503} no.~1, (2021) 920--937},
  \href{http://arxiv.org/abs/2007.02958}{{\ttfamily arXiv:2007.02958
  [astro-ph.GA]}}.

\bibitem{Hayashi:2020syu}
K.~Hayashi, M.~Ibe, S.~Kobayashi, Y.~Nakayama, and S.~Shirai, ``{Probing dark
  matter self-interaction with ultrafaint dwarf galaxies},''
  \href{http://dx.doi.org/10.1103/PhysRevD.103.023017}{{\em Phys. Rev. D}
  {\bfseries 103} no.~2, (2021) 023017},
  \href{http://arxiv.org/abs/2008.02529}{{\ttfamily arXiv:2008.02529
  [astro-ph.CO]}}.

\bibitem{Hisano:2002fk}
J.~Hisano, S.~Matsumoto, and M.~M. Nojiri, ``{Unitarity and higher order
  corrections in neutralino dark matter annihilation into two photons},''
  \href{http://dx.doi.org/10.1103/PhysRevD.67.075014}{{\em Phys. Rev. D}
  {\bfseries 67} (2003) 075014},
  \href{http://arxiv.org/abs/hep-ph/0212022}{{\ttfamily arXiv:hep-ph/0212022}}.

\bibitem{Hisano:2003ec}
J.~Hisano, S.~Matsumoto, and M.~M. Nojiri, ``{Explosive dark matter
  annihilation},'' \href{http://dx.doi.org/10.1103/PhysRevLett.92.031303}{{\em
  Phys. Rev. Lett.} {\bfseries 92} (2004) 031303},
  \href{http://arxiv.org/abs/hep-ph/0307216}{{\ttfamily arXiv:hep-ph/0307216}}.

\bibitem{Hisano:2004ds}
J.~Hisano, S.~Matsumoto, M.~M. Nojiri, and O.~Saito, ``{Non-perturbative effect
  on dark matter annihilation and gamma ray signature from galactic center},''
  \href{http://dx.doi.org/10.1103/PhysRevD.71.063528}{{\em Phys. Rev. D}
  {\bfseries 71} (2005) 063528},
  \href{http://arxiv.org/abs/hep-ph/0412403}{{\ttfamily arXiv:hep-ph/0412403}}.

\bibitem{Cirelli:2007xd}
M.~Cirelli, A.~Strumia, and M.~Tamburini, ``{Cosmology and Astrophysics of
  Minimal Dark Matter},''
  \href{http://dx.doi.org/10.1016/j.nuclphysb.2007.07.023}{{\em Nucl. Phys. B}
  {\bfseries 787} (2007) 152--175},
  \href{http://arxiv.org/abs/0706.4071}{{\ttfamily arXiv:0706.4071 [hep-ph]}}.

\bibitem{Arkani-Hamed:2008hhe}
N.~Arkani-Hamed, D.~P. Finkbeiner, T.~R. Slatyer, and N.~Weiner, ``{A Theory of
  Dark Matter},'' \href{http://dx.doi.org/10.1103/PhysRevD.79.015014}{{\em
  Phys. Rev. D} {\bfseries 79} (2009) 015014},
  \href{http://arxiv.org/abs/0810.0713}{{\ttfamily arXiv:0810.0713 [hep-ph]}}.

\bibitem{Sommerfeld:1931}
A.~Sommerfeld, ``{\"U}ber die beugung und bremsung der elektronen,''
  \href{http://dx.doi.org/https://doi.org/10.1002/andp.19314030302}{{\em
  Annalen der Physik} {\bfseries 403} no.~3, (1931) 257--330}.
  \url{https://onlinelibrary.wiley.com/doi/abs/10.1002/andp.19314030302}.

\bibitem{PAMELA:2008gwm}
{\bfseries PAMELA} Collaboration, O.~Adriani {\em et~al.}, ``{An anomalous
  positron abundance in cosmic rays with energies 1.5-100 GeV},''
  \href{http://dx.doi.org/10.1038/nature07942}{{\em Nature} {\bfseries 458}
  (2009) 607--609}, \href{http://arxiv.org/abs/0810.4995}{{\ttfamily
  arXiv:0810.4995 [astro-ph]}}.

\bibitem{Fermi-LAT:2011baq}
{\bfseries Fermi-LAT} Collaboration, M.~Ackermann {\em et~al.}, ``{Measurement
  of separate cosmic-ray electron and positron spectra with the Fermi Large
  Area Telescope},''
  \href{http://dx.doi.org/10.1103/PhysRevLett.108.011103}{{\em Phys. Rev.
  Lett.} {\bfseries 108} (2012) 011103},
  \href{http://arxiv.org/abs/1109.0521}{{\ttfamily arXiv:1109.0521
  [astro-ph.HE]}}.

\bibitem{AMS:2019rhg}
{\bfseries AMS} Collaboration, M.~Aguilar {\em et~al.}, ``{Towards
  Understanding the Origin of Cosmic-Ray Positrons},''
  \href{http://dx.doi.org/10.1103/PhysRevLett.122.041102}{{\em Phys. Rev.
  Lett.} {\bfseries 122} no.~4, (2019) 041102}.

\bibitem{Bringmann:2016din}
T.~Bringmann, F.~Kahlhoefer, K.~Schmidt-Hoberg, and P.~Walia, ``{Strong
  constraints on self-interacting dark matter with light mediators},''
  \href{http://dx.doi.org/10.1103/PhysRevLett.118.141802}{{\em Phys. Rev.
  Lett.} {\bfseries 118} no.~14, (2017) 141802},
  \href{http://arxiv.org/abs/1612.00845}{{\ttfamily arXiv:1612.00845
  [hep-ph]}}.

\bibitem{Kamada:2020buc}
A.~Kamada, H.~J. Kim, and T.~Kuwahara, ``{Maximally self-interacting dark
  matter: models and predictions},''
  \href{http://dx.doi.org/10.1007/JHEP12(2020)202}{{\em JHEP} {\bfseries 12}
  (2020) 202}, \href{http://arxiv.org/abs/2007.15522}{{\ttfamily
  arXiv:2007.15522 [hep-ph]}}.

\bibitem{Watson:1952ji}
K.~M. Watson, ``{The Effect of final state interactions on reaction
  cross-sections},'' \href{http://dx.doi.org/10.1103/PhysRev.88.1163}{{\em
  Phys. Rev.} {\bfseries 88} (1952) 1163--1171}.

\bibitem{Iengo:2009ni}
R.~Iengo, ``{Sommerfeld enhancement: General results from field theory
  diagrams},'' \href{http://dx.doi.org/10.1088/1126-6708/2009/05/024}{{\em
  JHEP} {\bfseries 05} (2009) 024},
  \href{http://arxiv.org/abs/0902.0688}{{\ttfamily arXiv:0902.0688 [hep-ph]}}.

\bibitem{Cassel:2009wt}
S.~Cassel, ``{Sommerfeld factor for arbitrary partial wave processes},''
  \href{http://dx.doi.org/10.1088/0954-3899/37/10/105009}{{\em J. Phys. G}
  {\bfseries 37} (2010) 105009},
  \href{http://arxiv.org/abs/0903.5307}{{\ttfamily arXiv:0903.5307 [hep-ph]}}.

\bibitem{Tulin:2013teo}
S.~Tulin, H.-B. Yu, and K.~M. Zurek, ``{Beyond Collisionless Dark Matter:
  Particle Physics Dynamics for Dark Matter Halo Structure},''
  \href{http://dx.doi.org/10.1103/PhysRevD.87.115007}{{\em Phys. Rev. D}
  {\bfseries 87} no.~11, (2013) 115007},
  \href{http://arxiv.org/abs/1302.3898}{{\ttfamily arXiv:1302.3898 [hep-ph]}}.

\bibitem{muschelivsvili1953singular}
N.~I. Muscheli{\v{s}}vili and J.~R.~M. Radok, {\em Singular integral equations:
  boundary problems of function theory and their application to mathematical
  physics}.
\newblock Wolters-Noordhoff publishing, 1953.

\bibitem{Omnes:1958hv}
R.~Omnes, ``{On the Solution of certain singular integral equations of quantum
  field theory},'' \href{http://dx.doi.org/10.1007/BF02747746}{{\em Nuovo Cim.}
  {\bfseries 8} (1958) 316--326}.

\bibitem{Levinson:407337}
N.~Levinson, ``{On the uniqueness of tne potential in a Schr{\"o}dinger
  equation for a given asymptotic phase},'' {\em Danske Vid. Selsk. Mat.-Fys.
  Medd.} {\bfseries 25} no.~9, (1949) 29.
  \url{http://cds.cern.ch/record/407337}.

\bibitem{Bethe:1949yr}
H.~A. Bethe, ``{Theory of the Effective Range in Nuclear Scattering},''
  \href{http://dx.doi.org/10.1103/PhysRev.76.38}{{\em Phys. Rev.} {\bfseries
  76} (1949) 38--50}.

\bibitem{Blatt:1949zz}
J.~M. Blatt and J.~D. Jackson, ``{On the Interpretation of Neutron-Proton
  Scattering Data by the Schwinger Variational Method},''
  \href{http://dx.doi.org/10.1103/PhysRev.76.18}{{\em Phys. Rev.} {\bfseries
  76} (1949) 18--37}.

\bibitem{Chu:2019awd}
X.~Chu, C.~Garcia-Cely, and H.~Murayama, ``{A Practical and Consistent
  Parametrization of Dark Matter Self-Interactions},''
  \href{http://dx.doi.org/10.1088/1475-7516/2020/06/043}{{\em JCAP} {\bfseries
  06} (2020) 043}, \href{http://arxiv.org/abs/1908.06067}{{\ttfamily
  arXiv:1908.06067 [hep-ph]}}.

\bibitem{Kang:2020afi}
Y.-J. Kang and H.~M. Lee, ``{Effective theory for self-interacting dark matter
  and massive spin-2 mediators},''
  \href{http://dx.doi.org/10.1088/1361-6471/abe529}{{\em J. Phys. G} {\bfseries
  48} no.~4, (2021) 045002}, \href{http://arxiv.org/abs/2003.09290}{{\ttfamily
  arXiv:2003.09290 [hep-ph]}}.

\bibitem{Cline:2022leq}
J.~M. Cline and C.~Perron, ``{Self-interacting dark baryons},''
  \href{http://dx.doi.org/10.1103/PhysRevD.106.083514}{{\em Phys. Rev. D}
  {\bfseries 106} no.~8, (2022) 083514},
  \href{http://arxiv.org/abs/2204.00033}{{\ttfamily arXiv:2204.00033
  [hep-ph]}}.

\bibitem{Kondo:2022lgg}
D.~Kondo, R.~McGehee, T.~Melia, and H.~Murayama, ``{Linear sigma dark
  matter},'' \href{http://dx.doi.org/10.1007/JHEP09(2022)041}{{\em JHEP}
  {\bfseries 09} (2022) 041}, \href{http://arxiv.org/abs/2205.08088}{{\ttfamily
  arXiv:2205.08088 [hep-ph]}}.

\bibitem{Chu:2018fzy}
X.~Chu, C.~Garcia-Cely, and H.~Murayama, ``{Velocity Dependence from Resonant
  Self-Interacting Dark Matter},''
  \href{http://dx.doi.org/10.1103/PhysRevLett.122.071103}{{\em Phys. Rev.
  Lett.} {\bfseries 122} no.~7, (2019) 071103},
  \href{http://arxiv.org/abs/1810.04709}{{\ttfamily arXiv:1810.04709
  [hep-ph]}}.

\bibitem{Blum:2016nrz}
K.~Blum, R.~Sato, and T.~R. Slatyer, ``{Self-consistent Calculation of the
  Sommerfeld Enhancement},''
  \href{http://dx.doi.org/10.1088/1475-7516/2016/06/021}{{\em JCAP} {\bfseries
  06} (2016) 021}, \href{http://arxiv.org/abs/1603.01383}{{\ttfamily
  arXiv:1603.01383 [hep-ph]}}.

\bibitem{Greene:1976a}
R.~L. Greene and C.~Aldrich, ``Variational wave functions for a screened
  coulomb potential,'' \href{http://dx.doi.org/10.1103/PhysRevA.14.2363}{{\em
  Phys. Rev. A} {\bfseries 14} (Dec, 1976) 2363--2366}.
  \url{https://link.aps.org/doi/10.1103/PhysRevA.14.2363}.

\bibitem{Slatyer:2009vg}
T.~R. Slatyer, ``{The Sommerfeld enhancement for dark matter with an excited
  state},'' \href{http://dx.doi.org/10.1088/1475-7516/2010/02/028}{{\em JCAP}
  {\bfseries 02} (2010) 028}, \href{http://arxiv.org/abs/0910.5713}{{\ttfamily
  arXiv:0910.5713 [hep-ph]}}.

\bibitem{Feng:2010zp}
J.~L. Feng, M.~Kaplinghat, and H.-B. Yu, ``{Sommerfeld Enhancements for Thermal
  Relic Dark Matter},''
  \href{http://dx.doi.org/10.1103/PhysRevD.82.083525}{{\em Phys. Rev. D}
  {\bfseries 82} (2010) 083525},
  \href{http://arxiv.org/abs/1005.4678}{{\ttfamily arXiv:1005.4678 [hep-ph]}}.

\bibitem{Wellner:1964}
M.~{Wellner}, ``{Levinson's Theorem (an Elementary Derivation)},''
  \href{http://dx.doi.org/10.1119/1.1969857}{{\em American Journal of Physics}
  {\bfseries 32} no.~10, (Oct., 1964) 787--789}.

\bibitem{Weinberg2015lectures}
S.~Weinberg, {\em Lectures on quantum mechanics}.
\newblock Cambridge University Press, 2015.

\end{thebibliography}\endgroup

\end{document}